\documentclass[aps,pre,showpacs,showkeys,floatfix,superscriptaddress]{revtex4-1}
\usepackage{amssymb}
\usepackage{amsmath}
\usepackage{amscd}
\usepackage{graphicx}
\usepackage{amsbsy}
\usepackage{color}
\usepackage{bm}
\begin{document}
\title{ From square-well to Janus: Improved algorithm for integral equation theory and comparison with thermodynamic perturbation theory within the Kern-Frenkel model}
\author{Achille Giacometti}
\affiliation{Dipartimento di Scienze Molecolari e Nanosistemi, Universit\`a Ca' Foscari Venezia,
Calle Larga S. Marta DD2137, I-30123 Venezia, Italy}
\email{achille.giacometti@unive.it}
\author{Christoph G\"ogelein}
\affiliation{Max-Planck-Institute for Dynamics and Self-Organization, G\"ottingen, Germany}
\email{christoph.goegelein@ds.mpg.de}
\author{Fred Lado}
\affiliation{Department of Physics, North Carolina State University, Raleigh, North Carolina 
27695-8202}
\email{lado@ncsu.edu}
\author{Francesco Sciortino}
\affiliation{{Dipartimento di Fisica and CNR-SOFT, 
Universit\`a di Roma {\em La Sapienza}, Piazzale A. Moro  2, 00185 Roma, Italy}}
\author{Silvano Ferrari}
\affiliation{Institut f{\"u}r Theoretische Physik and Center for Computational Materials Science, Technische Universit{\"a}t Wien, Wiedner Hauptstra{\ss}e 8-10/136, A-1040 Wien, Austria
}
\email{silvano.ferrari@tuwien.ac.at}
\author{Giorgio Pastore}
\affiliation{Dipartimento di Fisica  dell' Universit\`a di Trieste and CNR-IOM, 
Strada Costiera 11, 34151 Trieste, Italy}
\email{pastore@ts.infn.it}

\date{\today}
\begin{abstract}
Building upon past work on the phase diagram of Janus fluids [Sciortino et al., Phys. Rev. Lett. \textbf{103}, 237801 (2009)], we
perform a detailed study of integral equation theory of the Kern-Frenkel potential with coverage that is tuned from the
isotropic square-well fluid to the Janus limit. An improved algorithm for the reference hypernetted-chain (RHNC) equation for this
problem is implemented that significantly extends the range of applicability of RHNC. Results for both  structure and 
thermodynamics are presented and compared with numerical simulations. Unlike previous attempts, this algorithm is shown to
be stable down to the Janus limit, thus paving the way for analyzing the frustration mechanism characteristic of the gas-liquid
transition in the Janus system. The results are also compared with Barker-Henderson  thermodynamic perturbation theory 
on the same model. We then discuss the pros and cons of both approaches within a unified treatment. On balance, RHNC
integral equation theory, even with an isotropic hard-sphere reference system, is found to be a good compromise between accuracy of
the results, computational effort, and uniform quality to tackle self-assembly processes in patchy colloids of complex nature. 
Further improvement in RHNC however clearly requires an \emph{anisotropic} reference bridge function.
\end{abstract}

\keywords{Integral equation theory, Janus fluid, phase diagrams }
\maketitle
\section{Introduction}
\label{sec:intro}
Stimulated by recent advances in chemical syntheses of colloidal particles with different forms and functionalities, \cite{Walther09,Pawar10} theoretical approaches
have made significant progress in the last few years.
Patchy colloids \cite{Glotzer04,Glotzer07} in particular, having their surfaces decorated with different functionalities (e.g., solvophobic in opposition
to solvophilic moieties), appear to combine the possibility of obtaining a large number of targeted structures, on the one hand, along with the possibility of local rearrangements, on the other hand, that represent the optimal trade-off for engineering self-assembly processes at mesoscopic scales. \cite{Whitesides02}

While direct comparison of theory with experiment still relies heavily on extensive numerical simulations that constitute today the main theoretical tool, given their
 virtually exact  predictions, the heavy computational effort imposed by the anisotropic nature of patchy interactions (see e.g. Refs. \onlinecite{Sciortino09,Sciortino10}) has stimulated attempts to find approximate,  yet reliable, alternative methods that can provide semi-quantitative estimates within a modest amount of computer time.

Two of these methods with established roles in liquid state studies \cite{Gray84,Hansen86} are integral equation theory and thermodynamic perturbation theory. The main aim of integral equation theory is the computation of the pair correlation function, from which one can derive all thermodynamic and structural quantities.
In order to perform practical computations, one is forced to introduce here an approximation into the exact relation between pair potential and pair distribution function, i.e. selecting a closure equation. 
In thermodynamic perturbation theory, on the other hand, the free energy of the system can be computed as a perturbation series of terms, provided  the 
free energy and many-particle distribution functions  of a reference systems are known. Usually the expansion is approximated by the truncation of the infinite series to 
the few terms that can be evaluated.

In the present paper, we discuss the performances of both methods when applied to a particular model, the Kern-Frenkel potential \cite{Kern03,Chapman88} for patchy colloids, that has recently proven very useful within this anisotropic framework. Building upon previous work, \cite{Giacometti09a,Giacometti10,Giacometti09b,Gogelein12} we compare the performance of a specific integral equation closure, the reference hypernetted-chain (RHNC), \cite{Lado73,Rosenfeld79} and of a specific thermodynamic perturbation theory, devised by Barker and Henderson (TPT-BH), \cite{Barker67,Barker76} on the single-patch Kern-Frenkel potential. In the case of the RHNC integral equation, generalized for molecular fluids, \cite{Lado82a,Lado82b} we additionally present an improved algorithm allowing us to reach the limit of equal solvophobic-solvophilic composition, known as the Janus limit, that was not reachable with the original algorithm presented in Ref. \onlinecite{Giacometti09a}.

The remainder of the paper is organized as follows. In Sec. \ref{sec:ie}, we briefly recall the Kern-Frenkel model, while in Sec. \ref{sec:rhnc} and Sec. \ref{sec:TPT-BH} 
we review the application to this problem of the RHNC integral equation approach of Ref. \onlinecite{Giacometti09a} and the TPT-BH of Ref. \onlinecite{Gogelein12}.
The improved algorithm for RHNC is described in Sec. \ref{sec:improved} and a detailed comparison of the performance of the two methods in contrast to numerical simulations
is provided in Sec. \ref{sec:results}. Section \ref{sec:conclusions} completes the paper with some conclusions and perspectives.

\section{The Kern-Frenkel model}
\label{sec:ie}
The model for patchy interactions in colloids that we study here is due to Kern and Frenkel, \cite{Kern03} an elaboration of the original model by Chapman \textit{et al}. \cite{Chapman88} They consider a fluid
of hard spheres where the surface of each sphere is divided into two parts having square-well and hard-sphere character, the first
mimicking a solvophobic region, the second a solvophilic region, within an implicit solvent description.
Because of the azimuthal symmetry, the angular width of the solvophobic region is described  by a single polar angle $\theta_0$ that becomes equal to $\pi/2$ 
in the even-division case (the Janus limit).
 
\begin{figure}[htbp]
\begin{center}
\vskip0.5cm
\includegraphics[width=8cm]{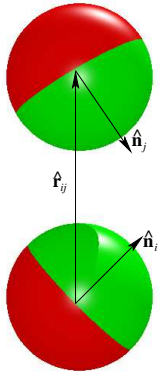} \\
\caption{The one-patch Kern-Frenkel model, where $\hat{\mathbf{r}}_{ij}$ is the direction joining the two centers and the orientations of the patches are specified by unit vectors $\hat{\mathbf{n}}_{i}$ and $\hat{\mathbf{n}}_{j}$. The present configuration depicts the Janus limit. 
\label{fig:fig1}}
\end{center}
\end{figure}

The positions of the $N$ particles in volume $V$ are given by a set of vectors $\mathbf{r}_i$, with $i=1,\ldots,N$, 
while the angular orientation of each square-well patch on a sphere surface is identified by unit vector  $\hat{\mathbf{n}}_{i}$. Finally, the direction connecting the centers of spheres $i$ and $j$ is characterized by unit vector $\hat{\mathbf{r}}_{ij}=(\mathbf{r}_j-\mathbf{r}_i)/ \vert \mathbf{r}_j-\mathbf{r}_i \vert$. Figure \ref{fig:fig1} depicts the situation in the case of the Janus limit.

Thus, two spheres of diameter $\sigma$ attract each other via a square-well potential of width $(\lambda-1) \sigma$ and depth $\epsilon$, if the directions 
of the patch on each sphere
are within a solid angle defined by $\theta_0$ and their relative distance lies within the range of the attractive well, 
and repel each other as hard spheres otherwise. As the system is still translationally invariant, the pair potential depends upon the difference
$\mathbf{r}_{ij}=\mathbf{r}_{j}-\mathbf{r}_{i}$, rather than $\mathbf{r}_{i}$ and $\mathbf{r}_{j}$ separately, and has the form \cite{Kern03,note1}
\begin{eqnarray}
\Phi\left(ij\right)\equiv \Phi\left(\mathbf{r}_{ij},\hat{\mathbf{n}}_{i},\hat{\mathbf{n}}_{j}\right) &=& \phi_{\text{HS}} \left(r_{ij}\right) +
\phi_{\text{SW}} \left(r_{ij}\right) \Psi\left(\mathbf{r}_{ij},\hat{\mathbf{n}}_{i},\hat{\mathbf{n}}_{j}\right),
\label{kf:eq1}
\end{eqnarray}
where $r_{ij}=\vert \mathbf{r}_{ij} \vert$. The first term in Eq. (\ref{kf:eq1}) is the hard-sphere (HS) contribution
\begin{equation}
\phi_{\text{HS}}\left(r\right)= \left\{ 
\begin{array}{ccc}
\infty,    &  &   0<r< \sigma    \\ 
0,          &  &   \sigma < r \ %
\end{array}%
\right.  \label{kf:eq2}
\end{equation}
while the second term can be factored into an isotropic square-well (SW) tail
\begin{equation}
\phi_{\text{SW}}\left(r\right)= \left\{ 
\begin{array}{ccc}
- \epsilon, &  &   \sigma<r< \lambda \sigma   \\ 
0,          &  &   \lambda \sigma < r \ %
\end{array}%
\right.  \label{kf:eq3}
\end{equation}
modulated by an angle-dependent factor
\begin{equation}
\Psi\left(\mathbf{r}_{ij},\hat{\mathbf{n}}_{i},\hat{\mathbf{n}}_{j}\right)= \left\{ 
\begin{array}{lllll}
1,  &   \text{if }     \hat{\mathbf{n}}_i \cdot \hat{\mathbf{r}}_{ij} \ge \cos \theta_0 & \text{and}  &
-\hat{\mathbf{n}}_j \cdot \hat{\mathbf{r}}_{ij} \ge \cos \theta_0, \\ 
0,  &   \text{otherwise.} & & & \
\end{array}
\right.  \label{kf:eq4}
\end{equation}
The unit vectors $\hat{\mathbf{n}}_{i}(\omega_{i})$  are defined by the spherical coordinates $\omega_i=(\theta_i,\varphi_i)$ in an arbitrarily oriented coordinate frame.
Here we will put $\beta\equiv\left(  k_{\mathrm{B}}T\right)  ^{-1}$, where $k_{\mathrm{B}}$ is Boltzmann's constant and $T$ is the absolute temperature, and introduce
the particle density $\rho=N/V$. We use reduced units for temperature, $T^*=k_{\mathrm{B}} T/\epsilon$, and density, $\rho^*=\rho \sigma^3$,
in the description of the thermodynamics. The above potential then ensures a proper bonding of the two particles depending upon the relative orientation and distance of the
attractive caps on each sphere.  

The square of the total coverage $\chi$ can be computed in terms of $\theta_0$ as
\begin{eqnarray}
\label{kf:eq5}
\chi^2 = \left\langle \Psi\left(\mathbf{r}_{ij},\hat{\mathbf{n}}_{i},\hat{\mathbf{n}}_{j}\right)
 \right \rangle_{\omega_i \omega_j} =\frac{1}{\left(4 \pi\right)^2}  \int d \omega_i d\omega_j
\Bigl[\Theta\left(\cos \theta_i-\cos \theta_0\right) \Theta\left(-\cos \theta_j-\cos \theta_0\right) \Bigr], 
\end{eqnarray}
where $\Theta(x)$ is the Heaviside step function, equal to $1$ if $x>0$ and $0$ if $x<0$, and where we have introduced the angular average
\begin{eqnarray}
\label{kf:eq6}
\left \langle \ldots \right \rangle_{\omega} &\equiv& \frac{1}{4 \pi} \int d \omega \ldots.
\end{eqnarray}
The integral can be readily evaluated to give \cite{Kern03}
\begin{eqnarray}
\label{kf:eq7}
\chi&=& \sin^2 \frac{\theta_0}{2}.
\end{eqnarray}
Knowledge of the exact result (\ref{kf:eq7}) of integral (\ref{kf:eq5}) is then exploited to optimize the discretization of the angular integration appearing in all successive integral equations, illustrated in the next Section.
\section{Molecular integral equation approach}
\label{sec:rhnc}
In the case of spherically symmetric potentials, the way to extract the thermophysical properties of a fluid has a long and venerable tradition in integral equation theory. Its central aim is the calculation of the pair distribution function $g(r)$, also typically computed in numerical simulations, from the pair potential  $\phi(r)$. It is useful as well to introduce the total correlation function 
$h(r)=g(r)-1$ and the so-called direct correlation function $c(r)$ defined through the Ornstein-Zernike (OZ) equation
\begin{equation}
\label{OZ}
h(r_{12}) = c(r_{12}) + \rho \int {\rm d} {\bf r_3} \, c( r_{13}) h(r_{32}).
\end{equation}
An exact, albeit formal, relation holds between such functions and the pair potential:
\begin{equation}
\label{exact}
g(r_{12}) = e^{ -\beta\phi(r_{12}) + h(r_{12}) - c(r_{12}) + B(r_{12}) },
\end{equation}
where the last term in the argument of the exponential is a (non-explicit) 
functional of the correlation function, generally called a bridge function for historical reasons. \cite{Green} All the existing approximations may be recast into the form of an approximate bridge function in Eq. (\ref{exact}), the so-called closure equation. Most current algorithms also invoke the use of the auxiliary function $\gamma(r)=h(r)-c(r)$, which is a continuous function even for discontinuous potentials such as hard spheres

The case of angle-dependent anisotropic potentials, although far more complex from an algorithmic point of view, follows essentially the same scheme.
It was devised in the frame of molecular fluids \cite{Gray84} and more recently 
adapted to the specific case of the Kern-Frenkel potential. \cite{Giacometti09b,Giacometti10} 
For completeness, the iterative procedure followed in Refs. \onlinecite{Giacometti09b,Giacometti10} is briefly reviewed below. 
\subsection{Iterative procedure}
\label{subsec:iterative}
Our notation in this section will closely follow that of Gray and Gubbins in Ref. \onlinecite{Gray84}, 
with only a $4 \pi$ prefactor difference; for instance,
$g(r;l_{1}l_{2}l)=4 \pi [g(r;l_{1}l_{2}l)]_{GG}$.
Starting with a reasonable guess for the set of coefficients $\gamma_{l_1 l_2 m}(r)$ in the axial $\mathbf{r}$-frame, 
where $\hat{\mathbf{z}}=\hat{\mathbf{r}}_{12}$, we use an
expansion in spherical harmonics to obtain $\gamma(12) \equiv  \gamma\left(r,\omega_1,\omega_2 \right) $ that in this frame depends only upon 
($r=r_{12},\omega_1,\omega_2$), 
\begin{eqnarray}
\label{rhnc:eq1}
\gamma(12)= 4\pi \sum_{l_1,l_2,m} \gamma_{l_1 l_2 m} 
\left(r\right) Y_{l_1m}\left(\omega_1\right) Y_{l_2\bar{m}}\left(\omega_2\right),
\end{eqnarray}
where $\bar{m}=-m$ and the $Y_{lm}\left(\omega\right)$ are spherical harmonics. Then we can use the closure relation
\begin{eqnarray}
\label{rhnc:eq2}
c\left(12\right)&=& \exp \left[-\beta \Phi\left(12\right)+\gamma\left(12\right)+
B\left(12\right)\right]-1-\gamma\left(12\right)
\end{eqnarray}
to obtain $c(12)$ that, in this frame, still depends only upon ($r,\omega_1,\omega_2$). The bridge function $B(12)$ in this expression must be approximated, giving rise to such distinct closures as Percus-Yevick (PY) and hypernetted-chain (HNC); see below for the reference HNC (RHNC)  closure used in this work. The inverse of an expansion like Eq. (\ref{rhnc:eq1})  is then used to compute the coefficients $c_{l_1 l_2 m}(r)$ within the same frame,
\begin{eqnarray}
\label{rhnc:eq3}
c_{l_{1} l_{2} m} \left(r\right)&=&\frac{1}{4\pi} \int d\omega_1d\omega_2\, c\left(r,\omega_1,\omega_2 \right) 
Y_{l_{1}m}^*\left(\omega_1\right)
Y_{l_{2}\bar{m}}^*\left(\omega_2\right) \nonumber \\
&\equiv&
4 \pi \left \langle c\left(r,\omega_1,\omega_2 \right) 
Y_{l_{1}m}^*\left(\omega_1\right)
Y_{l_{2}\bar{m}}^*\left(\omega_2\right) \right \rangle_{\omega_{1},\omega_{2}}. 
\end{eqnarray}
To carry out Fourier transforms and so deconvolute the molecular OZ equation, \cite{Gray84} we need to move at this point into an arbitrary space frame (often referred to as  laboratory-frame) by means of a Clebsch-Gordan (CG) transform,
\begin{eqnarray}
\label{rhnc:eq4}
c\left(r;l_{1} l_{2} l \right)&=& \left(\frac{4 \pi}{2l+1}\right)^{1/2}
\sum_{m} C \left(l_{1} l_{2} l ; m \bar{m} 0 \right) c_{l_{1} l_{2} m} \left(r\right),
\end{eqnarray}
where the $C \left(l_{1} l_{2} l ; m \bar{m} 0 \right)$ are Clebsch-Gordan coefficients. Fourier transforms then become Hankel transforms of the form
\begin{eqnarray}
\label{rhnc:eq5}
\widetilde{c}\left(k;l_{1} l_{2} l \right) &=& 4 \pi \mathrm{i}^{l}
\int_{0}^{\infty} dr~ r^2 c\left(r;l_{1} l_{2} l \right) j_{l} \left(k r\right),
\end{eqnarray}
where $j_l(x)$ is a spherical Bessel function of order $l$. We can then return to a specific frame, the axial $\mathbf{k}$-frame, where this time $\hat{\mathbf{z}}=
\hat{\mathbf{k}}$. This can be achieved by means of an inverse Clebsch-Gordan transform, 
\begin{eqnarray}
\label{rhnc:eq6}
\widetilde{c}_{l_{1} l_{2} m} \left(k\right)&=& \sum_{l} C \left(l_{1} l_{2} l ; m \bar{m} 0 \right) \left(\frac{2l+1}{4\pi}\right)^{1/2} 
\widetilde{c}\left(k;l_{1} l_{2} l\right).
\end{eqnarray}
Now one may use the Ornstein-Zernike equation in $k$ space, that in the axial $\mathbf{k}$-frame becomes
\begin{eqnarray}
\label{rhnc:eq7}
\widetilde{\gamma}_{l_{1} l_{2} m}\left(k\right) &=& \left(-1\right)^{m} \rho
\sum_{l_{3}=m}^{\infty} 
\left[\widetilde{\gamma}_{l_{1} l_{3} m}\left(k\right)+
\widetilde{c}_{l_{1} l_{3} m}\left(k\right) \right]
\widetilde{c}_{l_{3} l_{2} m}\left(k\right),
\end{eqnarray}
to obtain the new transform coefficients $\widetilde{\gamma}_{l_{1} l_{2} m}\left(k\right)$ by matrix operations. As before, one needs now to return to a more general space frame through a Clebsch-Gordan transform in Fourier space,
\begin{eqnarray}
\label{rhnc:eq8}
\widetilde{\gamma}\left(k;l_{1} l_{2} l \right)&=& \left(\frac{4 \pi}{2l+1}\right)^{1/2}
\sum_{m} C \left(l_{1} l_{2} l ; m \bar{m} 0 \right) \widetilde{\gamma}_{l_{1} l_{2} m} \left(k\right),
\end{eqnarray}
because this allows the return to direct space by means of an inverse Hankel transform,
\begin{eqnarray}
\label{rhnc:eq9}
\gamma\left(r;l_{1} l_{2} l \right) &=&
\frac{1}{2 \pi^2 \mathrm{i}^l} \int_{0}^{\infty} dk~ k^2 \,
\widetilde{\gamma}\left(k;l_{1} l_{2} l  \right) j_l \left(kr\right).
\end{eqnarray}
A final inverse Clebsch-Gordan transform then completes the return to the axial $\mathbf{r}$-frame we started with,
\begin{eqnarray}
\label{rhnc:eq10}
\gamma_{l_{1} l_{2} m} \left(r\right)&=& \sum_{l} C \left(l_{1} l_{2} l ; m \bar{m} 0 \right) \left(\frac{2l+1}{4\pi}\right)^{1/2}
\gamma\left(r;l_{1} l_{2} l\right),
\end{eqnarray}
and thus yields a new estimate of the starting coefficients $\gamma_{l_1 l_2 m}(r)$, in general different from the previous one. These steps are iterated
until consistency  between input and output coefficients $\gamma_{l_1 l_2 m}(r)$ is achieved. Table \ref{table:tab1} summarizes the procedure.
\begin{table*} 
{
\begin{equation*} 
\begin{CD}
c\left(r;l_{1}l_{2}l\right)
@>\text{Hankel transform Eq. (\ref{rhnc:eq5})}>>
\widetilde{c}\left(k;l_{1}l_{2}l\right)
@>\text{Inverse CG transform Eq. (\ref{rhnc:eq6})}>>
\widetilde{c}_{l_{1}l_{2}m}\left(k\right) \\
@AA\text{CG transform Eq. (\ref{rhnc:eq4})} A  @.   @VV\text{OZ equation Eq. (\ref{rhnc:eq7}) }   V\\
c_{l_{1}l_{2}m}\left(r\right)
@.      @.
\widetilde{\gamma}_{l_{1}l_{2}m}\left(k\right) \\     
@AA\text{Inverse expansion Eq. (\ref{rhnc:eq3})} A  @.    @VV\text{CG transform Eq. (\ref{rhnc:eq8})}V\\
c\left(r,\omega_1,\omega_2\right)
@.      @.
\widetilde{\gamma}\left(k;l_{1}l_{2}l\right) \\
@AA\text{Closure Eq. (\ref{rhnc:eq2})} A  @.    @VV\text{Inverse Hankel Eq. (\ref{rhnc:eq9})}V\\
\gamma\left(r,\omega_1,\omega_2\right)   
@.      @.
\gamma\left(r;l_{1} l_{2} l\right) \\
@AA\text{Expansion Eq. (\ref{rhnc:eq1})} A   @.   @VV\text{Inverse CG  Eq. (\ref{rhnc:eq10})}V\\
\left[\gamma_{l_{1} l_{2} m}\left(r\right) \right]_{\text{old}}   
@<<<    \text{Iterate}    @<<<  
\left[\gamma_{l_{1} l_{2} m}\left(r\right) \right]_{\text{new}}
\end{CD}
\end{equation*}
}
\caption{Schematic flow-chart for the solution of the OZ equation for the Kern-Frenkel angle-dependent potential.
See Section \ref{subsec:iterative} for a description of the scheme.}
\label{table:tab1}
\end{table*}
\subsection{The RHNC closure and free energy}
\label{subsec:closure}
Although the second equation in this scheme, Eq. (\ref{rhnc:eq2}), is formally exact, it involves the calculation of the bridge function $B(12)$ that in practice cannot be computed exactly, \cite{Hansen86} as remarked earlier, and so an approximate closure is needed.
Our approach is based on the RHNC approximation introduced in Ref. \onlinecite{Lado73} for spherical potentials and later extended to molecular fluids. \cite{Lado82a,Lado82b} Within this scheme, the closure equation takes on the assumed-known bridge function $B_0(12)$ of a particular reference system to replace the actual unknown bridge function $B(12)$ appearing in the exact closure. The goodness of the approximation clearly depends
upon the quality of the chosen bridge function for the reference system. In the present case, for want of a better option, this is taken to be the hard-sphere  model so that $B_0(12)=B_{\rm HS}(r_{12};\sigma_0)$, where $\sigma_0$ is the reference hard-sphere diameter. It has been demonstrated \cite{Rosenfeld79,Lado82} that internal thermodynamic consistency can be improved upon treating $\sigma_0$ as a variational parameter to be optimized. While the use of the hard-sphere bridge function is a natural assumption leading to a rather accurate approximation for spherically symmetric potentials, this is not as likely to be the case for a severely anisotropic potential such as the one-patch Kern-Frenkel model studied here. As we shall see below, this drawback is indeed confirmed by our findings, but better approximations for anisotropic potentials are not yet available.

Within the RHNC approximation, the excess free energy $F_{\rm ex}$ can be computed as \cite{Lado82b}
\begin{eqnarray}
\label{rhnc:eq11}
\frac{\beta F_{\rm ex}}{N} &=& 
\frac{\beta F_{1}}{N} +\frac{\beta F_{2}}{N}+\frac{\beta F_{3}}{N},
\end{eqnarray}
where
\begin{eqnarray}
\label{rhnc:eq12a}
\frac{\beta F_1}{N} &=& -\frac{1}{2} \rho \int d \mathbf{r}_{12} 
\left \langle \frac{1}{2} h^2\left(12\right)+h\left(12\right)-g\left(12\right) \ln \left
[ g\left(12\right) e^{\beta \Phi\left(12\right)} \right] \right \rangle_{\omega_{1} \omega_{2}}, \\
\label{rhnc:eq12b}
\frac{\beta F_2}{N}&=& - \frac{1}{2 \rho} \int \frac{d\mathbf{k}}{\left(2\pi\right)^3}
\sum_{m} \left \{ \ln \mathrm{Det} \left[ \mathbf{I} + \left(-1\right)^m \rho
\widetilde{\mathbf{h}}_m \left(k\right) \right] - \left(-1\right)^m
\rho\, \mathrm{Tr} \left[\widetilde{\mathbf{h}}_m \left(k\right) \right] \right \}, \\
\label{rhnc:eq12c}
\frac{\beta F_3}{N}&=& \frac{\beta F_3^{\rm ref}}{N}-\frac{1}{2} \rho \int d \mathbf{r}_{12} \left \langle \left[ g\left(12\right)-g_0\left(12\right) 
\right] 
B_0\left(12\right) \right 
\rangle_{\omega_{1} \omega_{2}}.
\end{eqnarray}
In Eq. (\ref{rhnc:eq12b}), $\widetilde{\mathbf{h}}_m(k)$ is a Hermitian matrix with elements $\widetilde{h}_{l_1 l_2 m}(k)$, $l_1,l_2 \ge m$, and $\mathbf{I}$ is the unit matrix. In Eq. (\ref{rhnc:eq12c}), $F_3$ directly expresses the RHNC approximation. Here $F_3^{\rm ref}$ is the reference system contribution, computed from the known free energy $F^{\rm ref}_{\rm ex}$ of the reference system as $F_3^{\rm ref}=F^{\rm ref}_{\rm ex}-F_1^{\rm ref}-F_2^{\rm ref}$, with $F_1^{\rm ref}$ and $F_2^{\rm ref}$ calculated as above but with reference system quantities. 

For the bridge function $B_0(12)=B_{\rm HS}(r_{12};\sigma_0)$ appearing in (\ref{rhnc:eq12c}), we use the Verlet-Weis-Henderson-Grundke parametrization, \cite{Verlet72,Henderson75} with the optimum hard sphere diameter $\sigma_0$ selected  according to a variational free energy minimization that yields the condition \cite{Lado82}
\begin{eqnarray}
\rho \int d \mathbf{r} \left [ g_{000}\left(r\right) - g_{\rm HS} 
\left(r;\sigma_0\right) \right] \sigma_0 \frac{\partial B_{\rm HS} \left(r;\sigma_0\right)}{\partial \sigma_0}&=&0.
\label{rhnc:eq13}
\end{eqnarray}
\subsection{Thermodynamics}
\label{subsec:thermodynamics}
The main strength of the RHNC closure hinges on the fact that, unlike most other closures, no further approximations are needed to obtain the free energy (as seen above) and other thermodynamic quantities. The pressure $P$ can be derived from a standard expression \cite{Gray84} as
\begin{eqnarray}
\label{rhnc:eq14}
P &=& \rho k_B T - \frac{1}{3V} \left \langle \sum_{i=1}^N \sum_{j>i}^N
r_{ij} \frac{\partial  \Phi\left( ij \right)}{\partial r_{ij}}
\right \rangle = \rho k_B T - \frac{1}{6} \rho^2 \int d \mathbf{r}_{12}
\left \langle g\left(12\right) r_{12} \frac{\partial  \Phi\left( 12 \right)}{\partial r_{12}}
\right \rangle_{\omega_{1} \omega_{2}}.
\end{eqnarray}
Introducing the cavity function $y(12)=g(12) e^{\beta \Phi(12)}$ and using the result
\begin{eqnarray}
\label{rhnc:eq15}
\frac{\partial}{\partial r} \left[ e^{-\beta \Phi(r,\omega_1,\omega_2)}\right] &=&
e^{\beta \epsilon \Psi\left(\omega_1,\omega_2\right)} \delta\left(r-\sigma \right)
- \left[ e^{\beta \epsilon \Psi\left(\omega_1,\omega_2\right)} -1 \right]
\delta\left(r-\lambda \sigma\right),
\end{eqnarray}
Eq. (\ref{rhnc:eq14}) becomes
\begin{eqnarray}
\label{rhnc:eq16}
\frac{\beta P}{\rho} &=& 1+ \frac{2}{3} \pi \rho \sigma^3 \left \{ 
\left \langle y\left(\sigma,\omega_1,\omega_2 \right) 
e^{\beta \epsilon \Psi\left(\omega_1,\omega_2\right)}\right \rangle_{\omega_1 \omega_2}
-  \lambda^3 \left \langle y\left(\lambda \sigma,\omega_1,\omega_2 \right) \left[
e^{\beta \epsilon \Psi\left(\omega_1,\omega_2\right)}-1 \right] \right \rangle_{\omega_1
\omega_2} \right \} \nonumber \\
&=& 1+ \frac{2}{3} \pi \rho \sigma^3 \left \{g_{000}\left(\sigma^{+}\right)+ \lambda^3 \left[g_{000}\left(\lambda \sigma^{+} \right) - g_{000}\left(\lambda \sigma^{-} \right)
\right] \right\}
\end{eqnarray}
which can be computed using Gaussian quadratures. Note that the second equality in Eq. (\ref{rhnc:eq14}) implies that the pressure depends upon the quality of
$g_{000}(r)$, the other components being irrelevant.  

The chemical potential $\mu$ can then be obtained from the exact thermodynamic relation
\begin{eqnarray}
\label{rhnc:eq17}
\beta \mu = \frac{\beta F}{N} + \frac{\beta P}{\rho},
\end{eqnarray}
with the ideal quantities given by $\beta F_{\rm id}/N = \ln (\rho \Lambda^3)-1$, $\beta P_{\rm id}/\rho=1$, $\beta \mu_{\rm id} = \ln (\rho \Lambda^3)$, where $\Lambda$ is the de Broglie wavelength.
\section{Improved Newton-Raphson algorithm}
\label{sec:improved}
The iteration cycle described in Section \ref{subsec:iterative}, wherein the output coefficients of one iteration directly become the input coefficients of the next, is known as Picard iteration. While obviously straightforward, it produces successive outputs that often converge only slowly or sometimes not at all, even for thermodynamic states that are known to exist. 
A  standard remedy is to construct the new input coefficients for the next iteration as a damping linear combination of the current input and output sets. \cite{Broyles60} We have implemented it in the efficient form proposed by  Ng \cite{Ng74} for generating a new input set of $\gamma_{l_1 l_2 m}(r)$ as an optimized linear superposition of the output sets from up to the previous four iterations.


But a more powerful procedure than such enhanced Picard cycles is available in the iterative application of Newton's well-known root-finding algorithm. In the present context, however, Newton's method, also known as the Newton-Raphson (NR) method, has the serious drawback of becoming so computationally intensive as to be prohibitive in practice, even for spherically symmetric models with just one coefficient. A clever meld of these two iteration techniques, producing a Newton-Raphson/Picard hybrid, was first proposed by Gillan \cite{Gillan79} for spherically symmetric models, using a small number of so-called roof functions to represent the ``coarse'' features of  $\gamma(r_i=i\Delta r)$ for NR processing. (Here $\Delta r$ is the grid interval  in the discrete $r$ space used in a numerical solution; the total number of grid points is $N_r$.)  Later,  Lab\'{i}k,  Malijevsk\'{y}, and Vo\v{n}ka (LMV) \cite{Labik85,Lomba89} suggested an elegant alternative based instead on the NR processing of a small number,  up to some cutoff $k_{\rm max}$, of  $\widetilde{\gamma}(k_i = i \Delta k)$ values, where $\Delta k$ is the grid interval in $k$ space. In this work, we have implemented the LMV hybrid, but for just the $\widetilde{\gamma}_{000}(k)$ coefficient, which makes the biggest contribution to $\widetilde{\gamma}(k,\omega_1,\omega_2)$, as explicitly illustrated by the results presented in Sec. \ref{subsec:RHNCvsMC}, while the other coefficients are treated by a standard Picard cycle. Not only does the algebra become unwieldy if more components are included in the NR iterations, but for the Kern-Frenkel potential there is no obvious basis for choosing {\em which} additional components to include.
We wish then to solve the one-component OZ equation (see Eq. (\ref{rhnc:eq7}))
\begin{equation}
\label{improved:eq1}
\widetilde{\gamma}_{000}(k_i) =  \rho \left[\widetilde{\gamma}_{000}(k_i)+\widetilde{c}_{000}(k_i) \right] \widetilde{c}_{000}(k_i),
\end{equation}
for $\widetilde{\gamma}_{000}(k_i)$ on the discrete $k_i$ grid, from $i=1$ to $i=n$, where $k_{\rm max}=n \Delta k$. Let $\widetilde{\Gamma}(k_i)$ be the desired solution, so that
\begin{equation}
\label{improved:eq2}
F[\widetilde{\Gamma}(k_i)] \equiv \widetilde{\Gamma}(k_i)-\frac{\rho \widetilde{c}_{000}^2(k_i)}{1-\rho \widetilde{c}_{000}(k_i)} = 0
\end{equation}
and $\widetilde{c}_{000}(k_i)$ is a function of all the $\widetilde{\Gamma}(k_j)$.  If  $\widetilde{\gamma}_{000}(k_i)$ is our current value for the unknown, then we need to find the correction  $\Delta \widetilde{\gamma}_{000}(k_i)$ such that 
$\widetilde{\Gamma}(k_i) = \widetilde{\gamma}_{000}(k_i)+\Delta \widetilde{\gamma}_{000}(k_i).$ 
This is accomplished in the NR root-finding method by setting
\begin{eqnarray}
\label{improved:eq3}
F[\widetilde{\gamma}_{000}(k_i)+\Delta \widetilde{\gamma}_{000}(k_i)] &\approx&  F[\widetilde{\gamma}_{000}(k_i)]
  +\sum_{j=1}^{n} C_{ij}\Delta \widetilde{\gamma}_{000}(k_j)=0, \\
C_{ij} &\equiv& \frac{\partial F[\widetilde{\gamma}_{000}(k_i)]}{\partial \widetilde{\gamma}_{000}(k_j)} \nonumber \\
          &=& \delta_{ij}-\left[ \frac{1}{(1-\rho \widetilde{c}_{000}(k_i))^2}-1 \right] \frac{\partial \widetilde{c}_{000}(k_i)}{\partial\widetilde{\gamma}_{000}(k_j)}.
\label{improved:eq4}
\end{eqnarray}
 Matrix inversion of Eq. (\ref{improved:eq3}) for the first $n$ points then produces the desired  corrections  $\Delta \widetilde{\gamma}_{000}(k_i)$.

Tracking the simplified one-component version of the Picard cycle in Section \ref{subsec:iterative}, $c_{000}(r_i)=g_{000}(r_i)-1-\gamma_{000}(r_i) \longrightarrow \widetilde{c}_{000}(k_j) \longrightarrow   \widetilde{\gamma}_{000}(k_j) \longrightarrow \gamma_{000}(r_i)$, leads to \cite{Labik85}
\begin{equation}
\label{improved:eq5}
\frac{\partial \widetilde{c}_{000}(k_i)}{\partial\widetilde{\gamma}_{000}(k_j)} = \frac{k_j}{k_i N_r} \sum_{l=1}^{N_r-1} h_{000}(r_l)
 \left\{ \cos \left[  l(i-j)\frac{\pi}{N_r} \right] -  \cos \left[ l(i+j)\frac{\pi}{N_r} \right] \right\}
\end{equation}
and completes the NR prescription. The discrete version \cite{lado67} of the reciprocal Fourier transforms requires that the intervals $\Delta r$ and $\Delta k$ 
satisfy $\Delta r \Delta k = \pi/N_r$. In the present calculations, we have used $\Delta r/\sigma= 0.02$, $N_r=1024$, and $n \approx 100$.
\section{Barker-Henderson thermodynamic perturbation theory}
\label{sec:TPT-BH}
Barker-Henderson perturbation theory \cite{Barker67,Henderson71,Barker76} hinges on the splitting of the Kern-Frenkel potential, Eq. (\ref{kf:eq1}), into
the hard-sphere contribution, Eq. (\ref{kf:eq2}), and the remaining ``perturbation''  term,
\begin{eqnarray}
\Phi_{I}\left(\mathbf{r}_{ij},\hat{\mathbf{n}}_{i},\hat{\mathbf{n}}_{j}\right) &\equiv& 
\phi_{\text{SW}} \left(r_{ij}\right) \Psi\left(\mathbf{r}_{ij},\hat{\mathbf{n}}_{i},\hat{\mathbf{n}}_{j}\right).
\label{tpt:eq1}
\end{eqnarray}
This allows the high-temperature expansion of the free energy as
\begin{eqnarray}
\label{tpt:eq2}
\frac{\beta \left(F - F_{\rm HS}\right)}{N} &=& f_1+f_2+ \ldots\,\mbox{,}
\end{eqnarray}
where $F_{\rm HS}$ is the free energy of the hard-sphere reference system, and where the first-order term, 
\begin{eqnarray}
\label{tpt:eq3}
f_1 &=& \frac{12 \eta}{\sigma^3} \int_\sigma^{\lambda\sigma} dr\, r^2 g_{\rm HS}\left(r\right) \left[ \beta \phi_{\text{SW}}(r) \right] 
\left \langle \Psi\left(12\right) \right \rangle_{\omega_1\omega_2}\,\mbox{,}
\end{eqnarray}
can be easily computed in terms of the radial distribution function $g_{\rm HS}(r)$ of the HS reference system; here $\eta = \pi\rho\sigma^3/6$ is the hard-sphere packing fraction. The second-order term is, on the contrary, a highly non-trivial calculation involving higher-order correlation functions.
An extension of the original Barker-Henderson alternative scheme yields the corresponding \textit{compressibility approximation} that reads \cite{Gogelein12}
\begin{eqnarray}
\label{tpt:eq4}
f_2 &=& -\frac{6 \eta}{\sigma^3}  \left(\frac{\partial \eta}{\partial P_0^{*}} \right)_T \int_\sigma^{\lambda\sigma} dr \, r^2
g_{\rm HS}\left(r\right) \left[ \beta \phi_{\text{SW}} \left(r\right) \right]^2
\left \langle \Psi^2 \left(12\right) \right \rangle_{\omega_1\omega_2}\,\mbox{,}
\end{eqnarray}
where $P_0^*=\beta P_0/\rho$ is the reduced pressure of the HS reference system in the Carnahan-Starling approximation. \cite{CS69} 
From here, pressure and chemical potential can be computed from the exact thermodynamic relations
\begin{eqnarray}
\label{tpt:eq5a}
\frac{\beta P}{\rho} &=& \eta \frac{\partial}{\partial \eta} \left(\frac{\beta F}{N} \right), \\
\label{tpt:eq5b}
\beta \mu &=& \frac{\partial}{\partial \eta} \left(\eta \frac{\beta F}{N} \right)\,\mbox{.}
\end{eqnarray}
\section{Results}
\label{sec:results}
\subsection{Pair distribution function}
\label{subsec:radial}
\begin{figure}[htbp]
\centering
\includegraphics[width=6.0in]{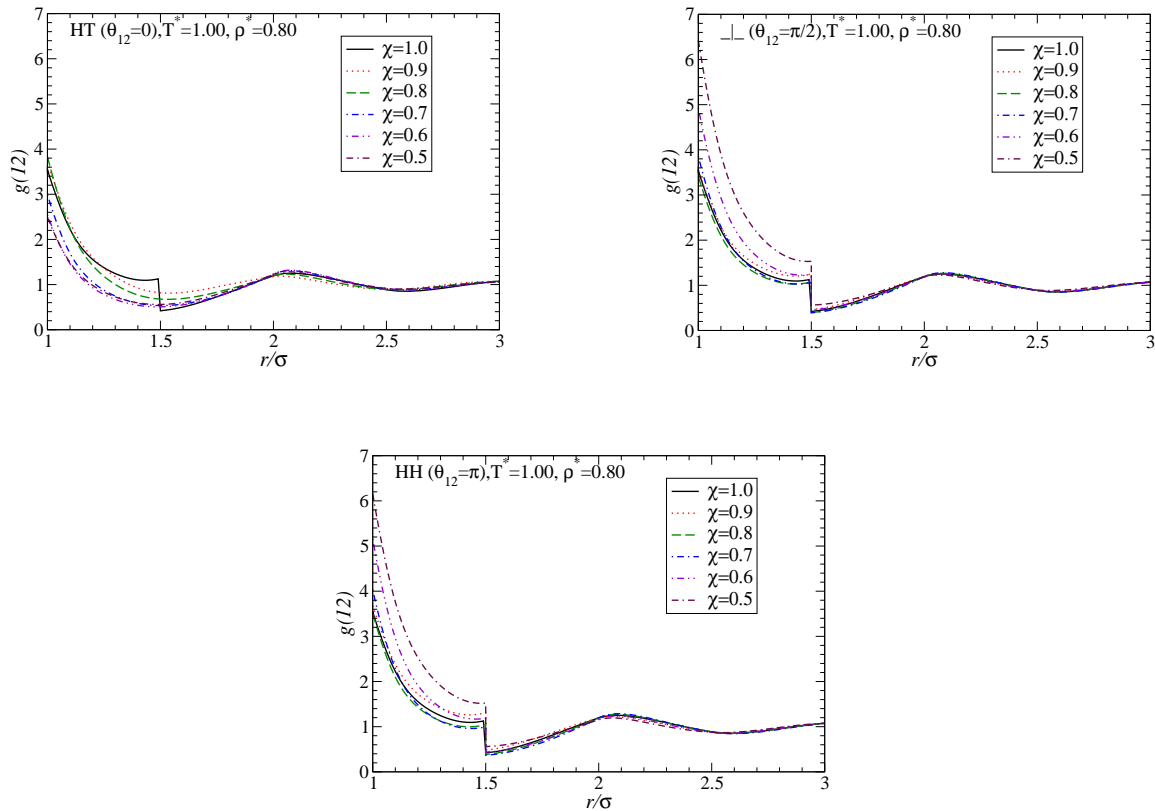} 
\caption{The $g(12)$ distribution function as a function of $r=\vert\mathbf{r}_{12} \vert$ for three orientations of the patches: 
HT,  $\hat{\mathbf{n}}_{1} \cdot \hat{\mathbf{n}}_{2} \equiv \cos \theta_{12} = 1$; 
$\perp$, $\hat{\mathbf{n}}_{1}\cdot \hat{\mathbf{n}}_{2} \equiv \cos \theta_{12} = 0$; 
HH,  $\hat{\mathbf{n}}_{1}\cdot \hat{\mathbf{n}}_{2} \equiv \cos \theta_{12} = -1$, 
and different coverages from $\chi=1.0$ (square-well) to $\chi=0.5$ (Janus).}
\label{fig:fig2}
\end{figure}

Unless otherwise stated, our results refer to  $\lambda=1.5$, as in Ref. \onlinecite{Giacometti09a}. Consider as initial state a reduced temperature $T^{*}=1.00$ for which the fluid is in a single phase at high density $\rho^{*}=0.8$
for all coverages examined here. We seek to determine the effect on the pair distribution function 
$g(12) \equiv g(r,\omega_1,\omega_2)$ of reducing the coverage $\chi$ for the given state point. This is reported in Fig. \ref{fig:fig2} for
three representative orientations: head-to-tail (HT), perpendicular ($\perp$), and head-to-head (HH), corresponding to angles 
$\theta_{12}\equiv\theta_2-\theta_1=0,\pi/2,\pi$ between the corresponding patch orientation vectors, respectively. 
(Similar plots were also considered in related systems, such as spherocylinders; see for instance Ref. \onlinecite{Martinez-Haya03})
\vskip0.5cm
Clearly, while for the HT $(\theta_{12}=0)$ case $g(12)$ is only mildly affected within the well, $\sigma<r<\lambda \sigma$,
both the $\perp$ $(\theta_{12}=\pi/2)$ and the HH $(\theta_{12}=\pi)$ pair distribution functions
display a significant increase close to the contact point $r=\sigma^{+}$.

On the other hand, the coexistence lines progressively shift to lower temperatures for decreasing coverages, as we will see, and hence a fixed state point
in the temperature-density plane is correspondingly moving relatively farther and farther from them, as coverage decreases.  

\begin{figure}[htbp] 
  \centering
   \includegraphics[width=6.0in]{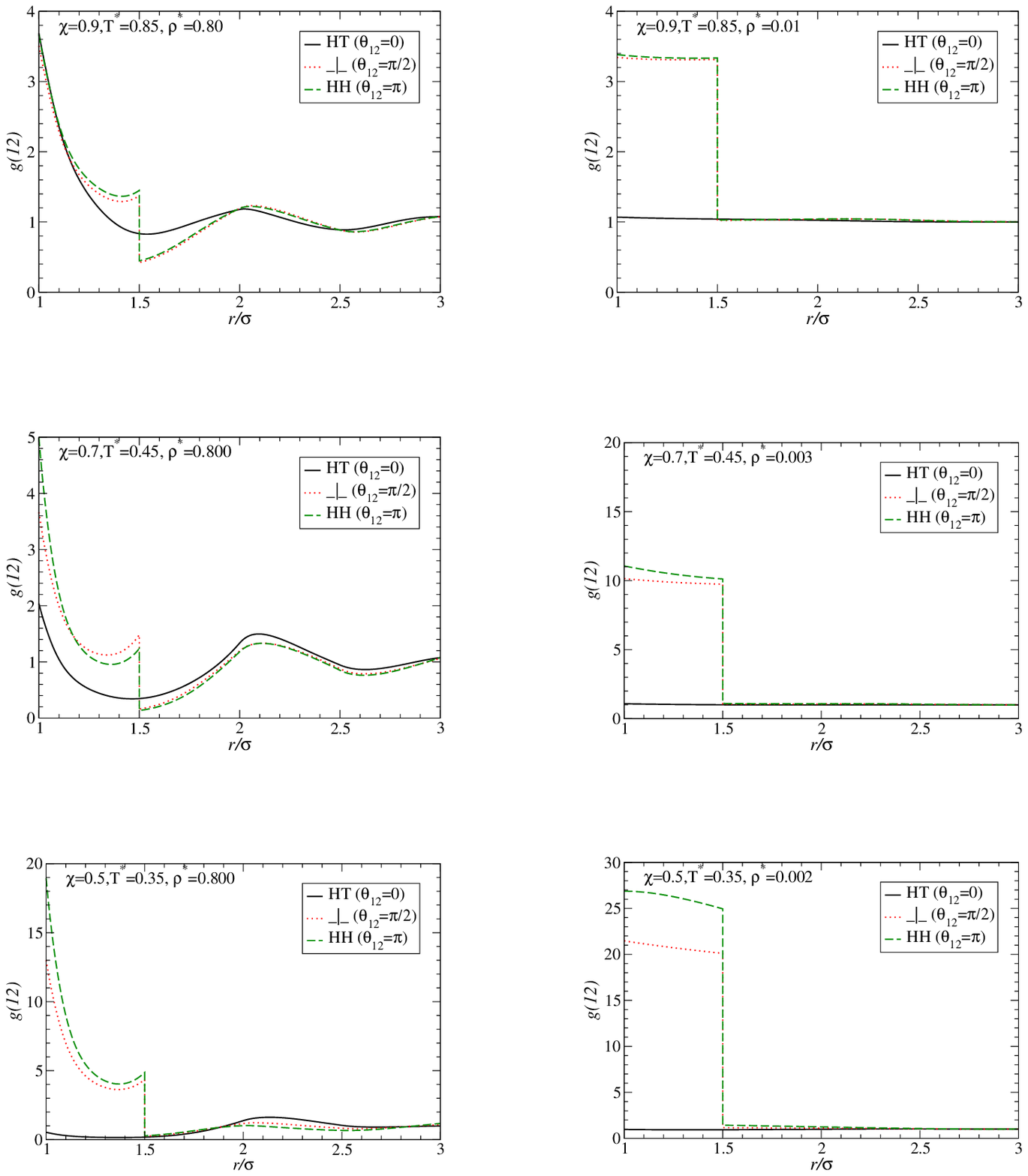} 
\caption{The $g(12)$  distribution function as a function of $r=\vert\mathbf{r}_{12} \vert$ for three orientations of the patches: 
HT,  $\hat{\mathbf{n}}_{1} \cdot \hat{\mathbf{n}}_{2} \equiv \cos \theta_{12} = 1$; 
$\perp$, $\hat{\mathbf{n}}_{1}\cdot \hat{\mathbf{n}}_{2} \equiv \cos \theta_{12} = 0$; 
HH,  $\hat{\mathbf{n}}_{1}\cdot \hat{\mathbf{n}}_{2} \equiv \cos \theta_{12} = -1$.
In all cases, we present results for the highest and lowest densities studied at the lowest temperatures achieved at each coverage.
From top to bottom, this corresponds to: 
$\chi=0.9$, $T^{*}=0.85$, $\rho^{*}=0.8$ (left), $\rho^{*}=0.010$ (right);
$\chi=0.7$, $T^{*}=0.45$, $\rho^{*}=0.8$ (left), $\rho^{*}=0.003$ (right);
$\chi=0.5$, $T^{*}=0.35$, $\rho^{*}=0.8$ (left), $\rho^{*}=0.002$ (right).}
\label{fig:fig3}
\end{figure}

In order to account for this and make different coverages comparable, we consider different state points 
that are comparably close to the gas-liquid coexistence lines. These are shown in Fig. \ref{fig:fig3} for decreasing coverage
from $\chi=0.9$ to $\chi=0.5$ and two specific state points, side by side, that have different temperatures for the different coverages. In each case, we have first considered
the largest computed density ($\rho^{*}=0.8$ for all coverages) and the corresponding lowest computed temperature (decreasing with decreasing
coverage). The panels on the left side of Fig. \ref{fig:fig3} correspond to state points expected to lie in the liquid phase at the respective coverages and are shown for decreasing coverage from top to bottom. The other set of chosen state points in the right-hand panels of Fig. \ref{fig:fig3} are all points
lying in the respective gas phases (low temperatures and low densities) and are depicted again for decreasing coverage from top to bottom. In all cases, three different curves are reported corresponding to the HT, $\perp$, and HH orientations of the two patches.

Consider first the high-density state points on the left. Few general features are readily apparent. In all cases, the HT curve
exhibits a hard-spheres behavior with no discontinuity at the well edge, $r=\lambda \sigma$, as expected from the definition of the 
Kern-Frenkel potential.

Note that the value of this $g(12)$ at contact, $r=\sigma^{+}$, decreases
as the coverage decreases, since it becomes less and less likely to find particles with the HT orientation of the patches as $\chi$ decreases (further
note the change in scale among different cases).
Conversely, both $\perp$ and HH curves exhibit the usual discontinuity at $r=\lambda \sigma$, indicating that they are involved in
bonding, with a progressive increase of the  $g(12)$ at contact, $r=\sigma^{+}$, as coverage decreases that is more marked in the HH
than in the $\perp$ case.

A rather interesting pattern emerges from the low-density plots of the right-hand panels. Those are the cases where one expects an increase in
micellization as coverage decreases.
This is indeed confirmed by the results. As coverage decreases, the general trend is a significant increase
of  $g(12)$ at contact, $r=\sigma^{+}$, the largest increase pertaining to the HH orientations, as expected. This clearly indicates
the formation of clusters (micelles or vesicles) with an increasing fraction of saturated bonds. In particular, in the Janus case ($\chi=0.5$)
the HT orientation gives a flat curve around $g(12)=1$, indicating an almost ideal behavior that reflects the almost complete absence of such orientations.
However, we have observed no significant discontinuity on passing from $\chi=0.6$ to $\chi=0.5$ coverages that would indicate anomalous behavior of the Janus
case. Therefore, RHNC is clearly not able to capture this effect with the present spherically-symmetric approximation of $B_0(12)$.

\subsection{Angular distributions}
\label{subsec:angular}
Complementary to previous cases, here we focus on the dependence of $g(12)$ on just the orientations of $\hat{\mathbf{n}}_{2}$ and $\hat{\mathbf{r}}_{12}$ relative to $\hat{\mathbf{n}}_{1}$ within the square-well region. The expansion in spherical harmonics $Y_{lm}(\omega)$ of $g(12)$ in an arbitrary space frame reads 
\begin{eqnarray}
\label{ang:eq1}
g\left(12\right) &=& \sum_{l_{1},l_{2}=0}^{\infty} \sum_{l=\vert l_1 - l_2 \vert}^{l_1+l_2} g^{l_{1} l_{2} l}\left(r\right)
\psi^{l_{1} l_{2} l}\left(\omega_1 \omega_2 \Omega\right),
\end{eqnarray}
where we have introduced the rotational invariants \cite{Gray84,Hansen86}
\begin{eqnarray}
\label{ang:eq2}
\psi^{l_{1} l_{2} l}\left(\omega_1 \omega_2 \Omega\right)&=& \sum_{m_{1}=-l_{1}}^{l_{1}}  \sum_{m_{2}=-l_{2}}^{l_{2}}
C\left(l_{1} l_{2} l; m_{1} m_{2} m_{1}+m_{2} \right)
Y_{l_{1}m_{1}}\left(\omega_{1}\right) Y_{l_{2}m_{2}}\left(\omega_{2}\right)
Y_{l, m_{1}+m_{2}}^{*}\left(\Omega\right).
\end{eqnarray}
In Ref. \onlinecite{Giacometti10}, it was shown that upon defining 
\begin{eqnarray}
\label{ang:eq6}
\bar{g}\left(l_1 l_2 l\right) &=& \frac{1}{4 \pi(\lambda-1) \sigma} \int_{\sigma}^{\lambda \sigma} dr g^{l_{1} l_{2} l}\left(r \right), \\
\bar{g}\left(\theta,\theta_2\right) &=& \frac{1}{(\lambda-1) \sigma} \int_{\sigma}^{\lambda \sigma} dr \left \langle g\left(12\right) \right \rangle_{\varphi_2 \varphi}. 
\end{eqnarray}
the resulting function of the polar coordinate $\theta$ of $\hat{\mathbf{r}}_{12}$ and the polar coordinate $\theta_2$ of the second patch reads
\begin{eqnarray}
\label{ang:eq7}
\bar{g}\left(\theta,\theta_2\right) &=& 
\sum_{l_{1},l_{2},l} \bar{g}\left(l_1 l_2 l\right)
\left[ \frac{\left(2l_1+1\right) \left(2l_2+1\right) \left(2l+1\right)}{4\pi} \right]^{1/2} C\left(l_{1} l_{2} l; 0 0 0 \right) P_{l_{2}}\left(\cos \theta_2\right) 
P_{l}\left(\cos \theta\right),
\end{eqnarray}
given that the $z$ axis is aligned with the patch of particle 1.

\begin{figure}[htbp] 
  \centering
   \includegraphics[width=6.0in]{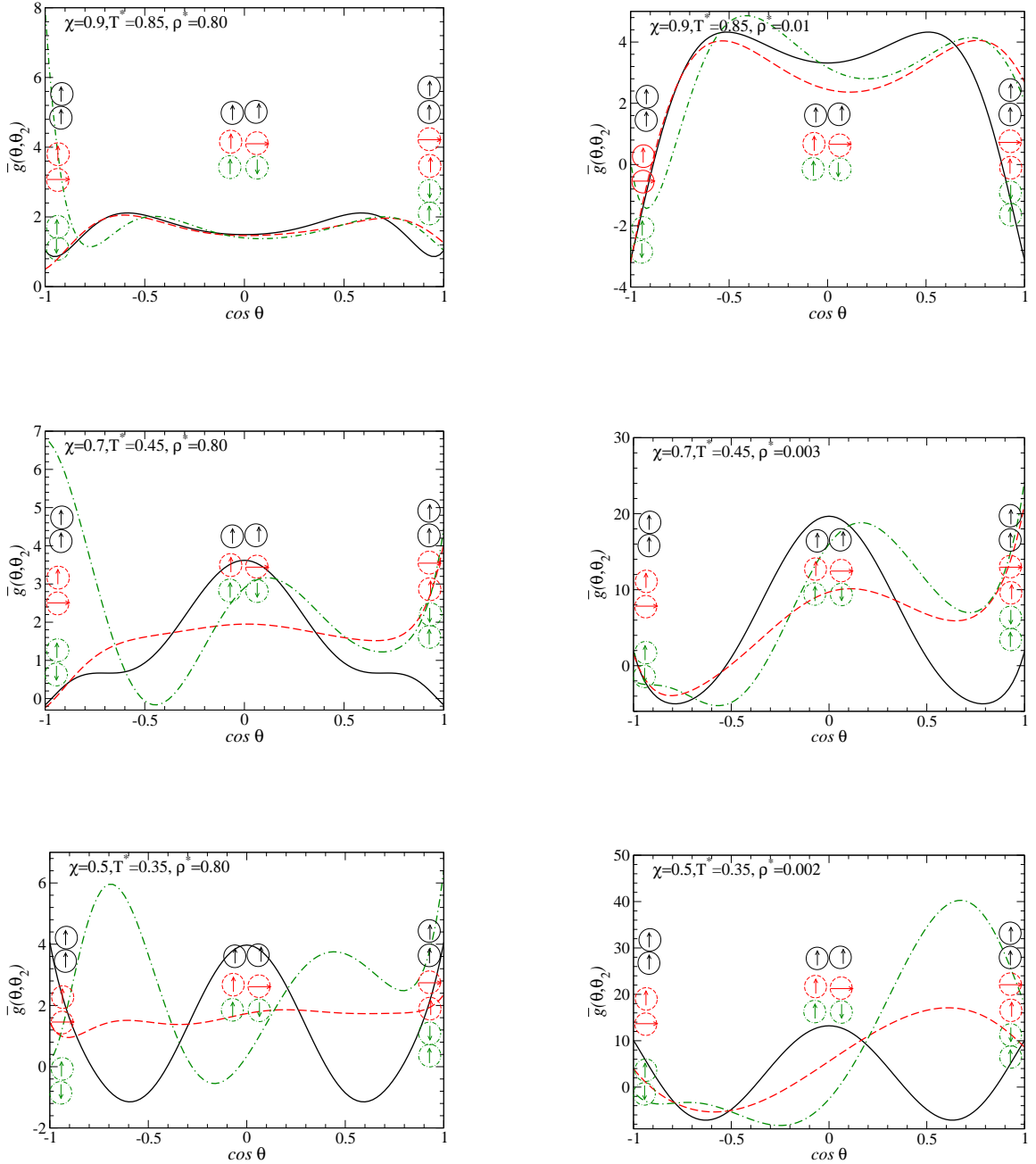} 
\caption{The $\overline{g}(\theta,\theta_2)$ angular distribution as a function of $\cos \theta$ for three orientations of 
patch $2$ ($\theta_{2} = 0$,  $\theta_{2} = \pi/2$, $\theta_{2} = \pi$) given that patch $1$ is pointing up ($\theta_{1}=0$):
In all cases, we present results for the highest and lowest densities studied at the lowest temperatures achieved at each coverage.
From top to bottom, this corresponds to: 
$\chi=0.9$, $T^{*}=0.85$, $\rho^{*}=0.8$ (left), $\rho^{*}=0.010$ (right);
$\chi=0.7$, $T^{*}=0.45$, $\rho^{*}=0.8$ (left), $\rho^{*}=0.003$ (right);
$\chi=0.5$, $T^{*}=0.35$, $\rho^{*}=0.8$ (left), $\rho^{*}=0.002$ (right).
The circular arrowed insets refer to the patch orientation, with $\theta_1=0$ (always up) and $\theta_{2}$ rotating.}
\label{fig:fig4}
\end{figure}

The behavior of $\bar{g}(\theta,\theta_2)$ as a function of $\cos \theta$ is reported in Fig. \ref{fig:fig4} for three different orientations
of the patches: HT ($\theta_1=0,\theta_2=0$), $\perp$ ($\theta_1=0,\theta_2=\pi/2$), HH ($\theta_1=0,\theta_2=\pi$), and different coverages
from $\chi=0.9$ to $\chi=0.5$. The same high and low densities state points used before have been considered here.
This identifies the preferential angular positions of the various different patch orientations.

Consider the high density state point first, depicted in the left-hand panels of Fig. \ref{fig:fig4} for decreasing coverages from top to bottom. State points are the same
discussed in Fig. \ref{fig:fig3}. For sufficiently
large patches ($\chi=0.9,0.8$, not shown here), the only significant peak in the distribution is observed for $\theta_2 \approx \pi$ and $\theta \approx \pi$. 
For such high coverages, HH alignments are uniformly distributed along all solid angles $0 \le \theta \le \pi$ (remember that there is
azimuthal symmetry), whereas HT alignment is preferentially found in the backward direction, $\theta \approx \pi$. 

The situation changes as the coverage decreases from $\chi=0.7$, with the development of further peaks for perpendicular orientation of the
patches ($\theta_1=0,\theta_2=\pi/2$) at $\theta \approx \pi/2$ and for head-to-head orientation of the patches ($\theta_1=0,\theta_2=\pi$) at  $\theta \approx 0$. The physical interpretation of these results is that, under high density and low temperature conditions, head-to-tail (HT) and head-to-head (HH) alignments of the patches are preferentially found for particles in the transversal direction, $\theta \approx \pi/2$, for low coverages ($\chi \le 0.7$).

Next, we consider the low density points reported in the right-hand panels of Fig. \ref{fig:fig4}, again for decreasing coverages from top to bottom. Unlike the previous case, we find a clear predominance of the HH antiparallel alignment in the forward direction ($\theta \approx 0$) and modulated layering for both
HT and $\perp$ patch orientations that become increasingly structured as coverage decreases. These results can be contrasted with the analogous results given in Ref. \onlinecite{Giacometti10} for the two-patch case and extend those given there for only high and low coverages.  The layering is a clear reflection of an increasing tendency to micellization, in agreement with numerical simulation results.

\subsection{Coefficients of rotational invariants}
\label{subsec:rotational}
In this section, we follow the the notations already introduced in our previous work. \cite{Giacometti10}
The coefficients of rotational invariants are
\begin{eqnarray}
\label{rotational:eq1}
g^{l_{1} l_{2} l}\left(r\right) &=& \frac{1}{4 \pi \rho r^2 N} \left \langle \sum_{i\ne j} \delta\left(r-r_{ij}\right) \Delta^{l_{1} l_{2} l} (12) \right \rangle,  
\end{eqnarray}
where the $\Delta^{l_{1} l_{2} l} (12)$ are rotational invariants. Here we have explicitly considered the first $10$ coefficients occurring in the multipole
expansion \cite{Gray84} that account up to quadrupole-quadrupole interactions. \cite{Stell81}

Explicit expressions for the first few are \cite{Stell81} 
\begin{eqnarray}
\label{rotational:eq2}
\Delta^{000}\left(12\right) &=&1, \\ \nonumber
\Delta^{110}\left(12\right) &=& 3\Delta\left(12\right)=3\,\hat{\mathbf{n}}_1 \cdot \hat{\mathbf{n}}_2, \\ \nonumber
\Delta^{112}\left(12\right)&=& \frac{3}{2} D\left(12\right)= \frac{3}{2} \left[3 \left(\hat{\mathbf{n}}_1 \cdot \hat{\mathbf{r}}_{12} \right)
\left(\hat{\mathbf{n}}_2 \cdot \hat{\mathbf{r}}_{12} \right) - \hat{\mathbf{n}}_1 \cdot \hat{\mathbf{n}}_2 \right], \\ \nonumber
\Delta^{220}\left(12 \right)&=& \frac{5}{2} E\left(12\right)=\frac{5}{2} \left[3\left( \hat{\mathbf{n}}_1 \cdot \hat{\mathbf{n}}_2\right)^2 -1\right].
\end{eqnarray} 
Other expressions can be found in Ref. \onlinecite{Stell81}. We note that $g^{l_{1} l_{2} l}(r)= 4 \pi g\left(r;l_{1} l_{2} l\right)$ used in past work \cite{Giacometti10} and further that $g^{l_{1} l_{2} l}(r)=h^{l_{1} l_{2} l}(r)+ \delta_{l_{1}0}\delta_{l_{2}0} \delta_{l0}$.

In Appendix \ref{app:appa}, we explicitly derive Eq. (\ref{rotational:eq1}) for two specific and representative cases.
Some of the coefficients have particularly interesting physical interpretations:   
the term $h^{110}(r)$ is the coefficient of ferroelectric correlation, the term $h^{112}(r)$ the coefficient of dipolar correlation, 
the term $h^{220}(r)$ the coefficient of nematic correlation, and so on.

\begin{figure}[htbp] 
  \centering
   \includegraphics[width=6.0in]{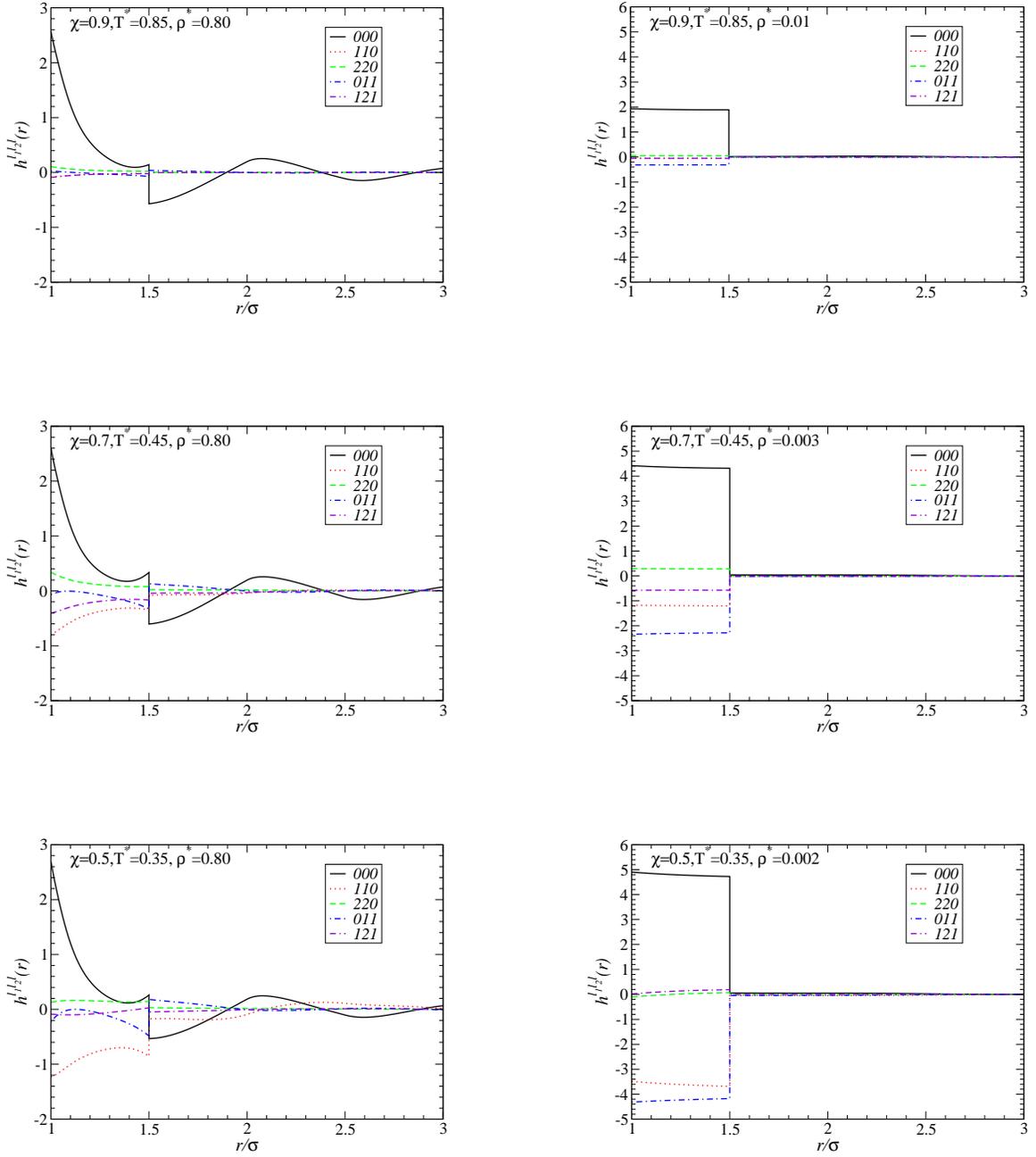} 
\caption{The $h^{l_1l_2l}(r)$ rotational invariants as a function of $r$ for several triplets. 
In all cases, we present results for the highest and lowest densities studied at the lowest temperatures achieved at each coverage.
From top to bottom, this corresponds to: 
$\chi=0.9$, $T^{*}=0.85$, $\rho^{*}=0.8$ (left), $\rho^{*}=0.010$ (right);
$\chi=0.7$, $T^{*}=0.45$, $\rho^{*}=0.8$ (left), $\rho^{*}=0.003$ (right);
$\chi=0.5$, $T^{*}=0.35$, $\rho^{*}=0.8$ (left), $\rho^{*}=0.002$ (right), as in Fig. \ref{fig:fig3}.
}
\label{fig:fig5}
\end{figure}

The results for these coefficients are reported in Fig. \ref{fig:fig5}, with the same ordering as before. Hence the left-hand panels show plots of the
high-density points and decreasing coverage, while the right-hand panels depict plots of the low-density points and again decreasing coverages.
Plots on the same side have been drawn to the same scale so that differences may be readily appreciated.

The high-density plots (left-hand panels) have hardly any dependence on the particular projection, as could have been guessed from the outset.
With $h^{000}(r)=g^{000}(r)-1$, we clearly find correlations (that is, non-vanishing coefficients) only within the
well, $\sigma < r < \lambda \sigma$, along with $h^{110}(r)$ and $h^{121}(r)$ negatively correlated, $h^{220}(r)$ positively correlated, and $h^{011}(r)$
almost uncorrelated. Similar behavior occurs for the low-density state points where, however, the correlation within the well is approximately constant, 
with $h^{011}(r)<h^{110}(r)<h^{121}(r)<0$, and $h^{220}(r)>0$. Note that in the last, Janus case ($\chi=0.5$), the $h^{220}(r)$ and $h^{121}(r)$ ordering
appear to be inverted, signaling an incomplete agreement with the other cases, likely due to an insufficient lowering of the temperature, in agreement with
previous findings of Sections \ref{subsec:radial}.

\begin{figure}[htbp] 
  \centering
   \includegraphics[width=6.0in]{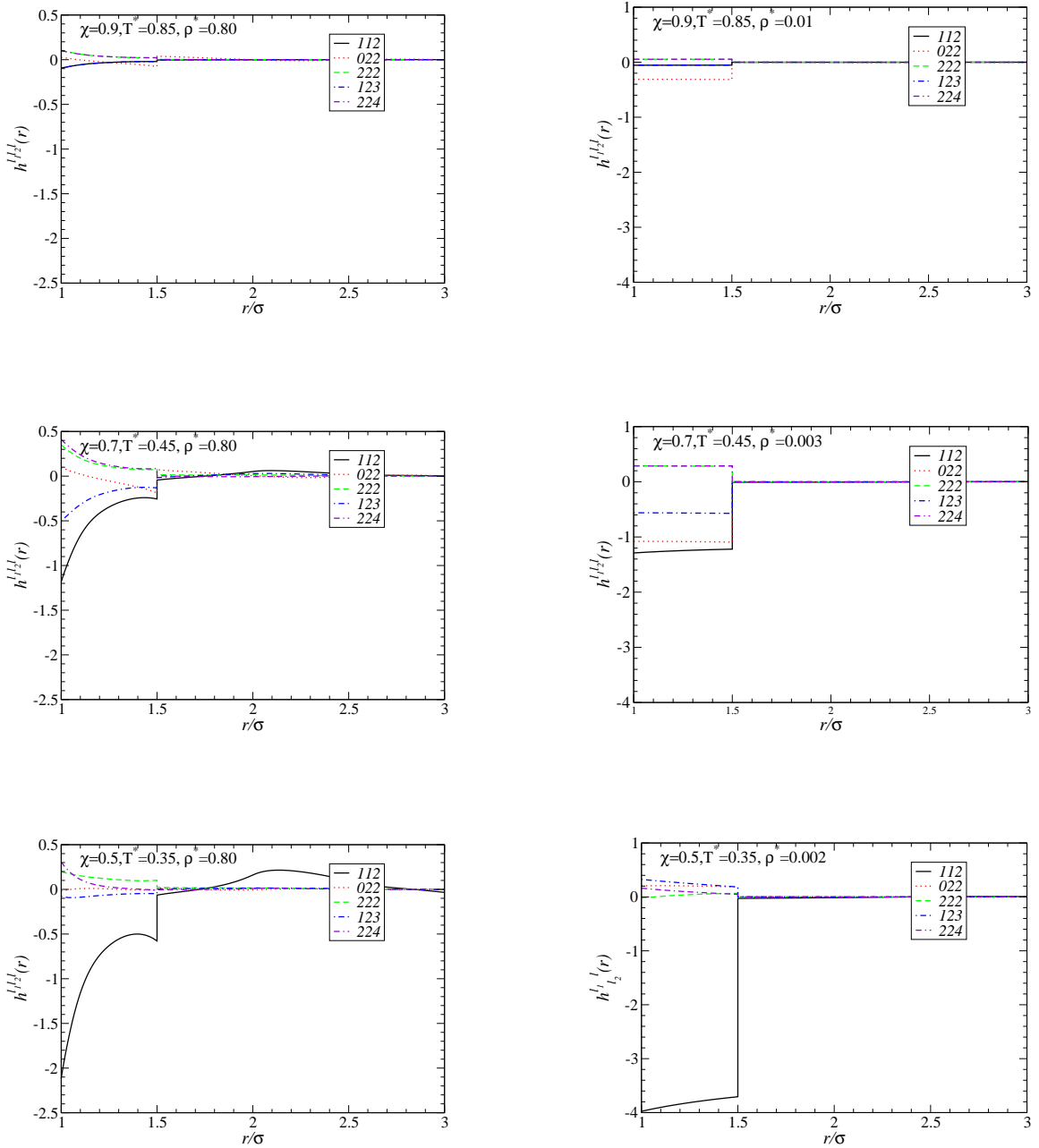} 
\caption{The $h^{l_1l_2l}(r)$ rotational invariants as a function of $r$ for several triplets. 
In all cases, we present results for the highest and lowest densities studied at the lowest temperatures achieved at each coverage.
From top to bottom, this corresponds to: 
$\chi=0.9$, $T^{*}=0.85$, $\rho^{*}=0.8$ (left), $\rho^{*}=0.010$ (right);
$\chi=0.7$, $T^{*}=0.45$, $\rho^{*}=0.8$ (left), $\rho^{*}=0.003$ (right);
$\chi=0.5$, $T^{*}=0.35$, $\rho^{*}=0.8$ (left), $\rho^{*}=0.002$ (right).}
\label{fig:fig6}
\end{figure}

Next we consider a second set of coefficients given by $h^{112}(r)$, $h^{022}(r)$, $h^{222}(r)$, $h^{123}(r)$, $h^{224}(r)$. These are reported in Fig. \ref{fig:fig6}
with the same distribution as before. Even in this case, all coefficients have non-vanishing values within the well and have thus been plotted to the same scale.
Again, the trend appears to be rather clear, with the coefficient $h^{112}(r)$ negative with
decreasing contact values for decreasing patch size, indicating an increasing anticorrelation in the respective orientations as coverage decreases;  $h^{123}(r)$
also has negative value, whereas all others coefficients present positive values indicating positive correlations. This is true for both high- and low-density states. 
\subsection{RHNC molecular reference angular components of radial distribution functions and MC results}
\label{subsec:RHNCvsMC}
In order to assess the structural results previously discussed, in this section we compare directly the RHNC molecular-frame spherical harmonic coefficients $g_{l_1l_2m}(r)$ and MC results.
\begin{figure}[htbp] 
  \centering
   \includegraphics[width=6.0in]{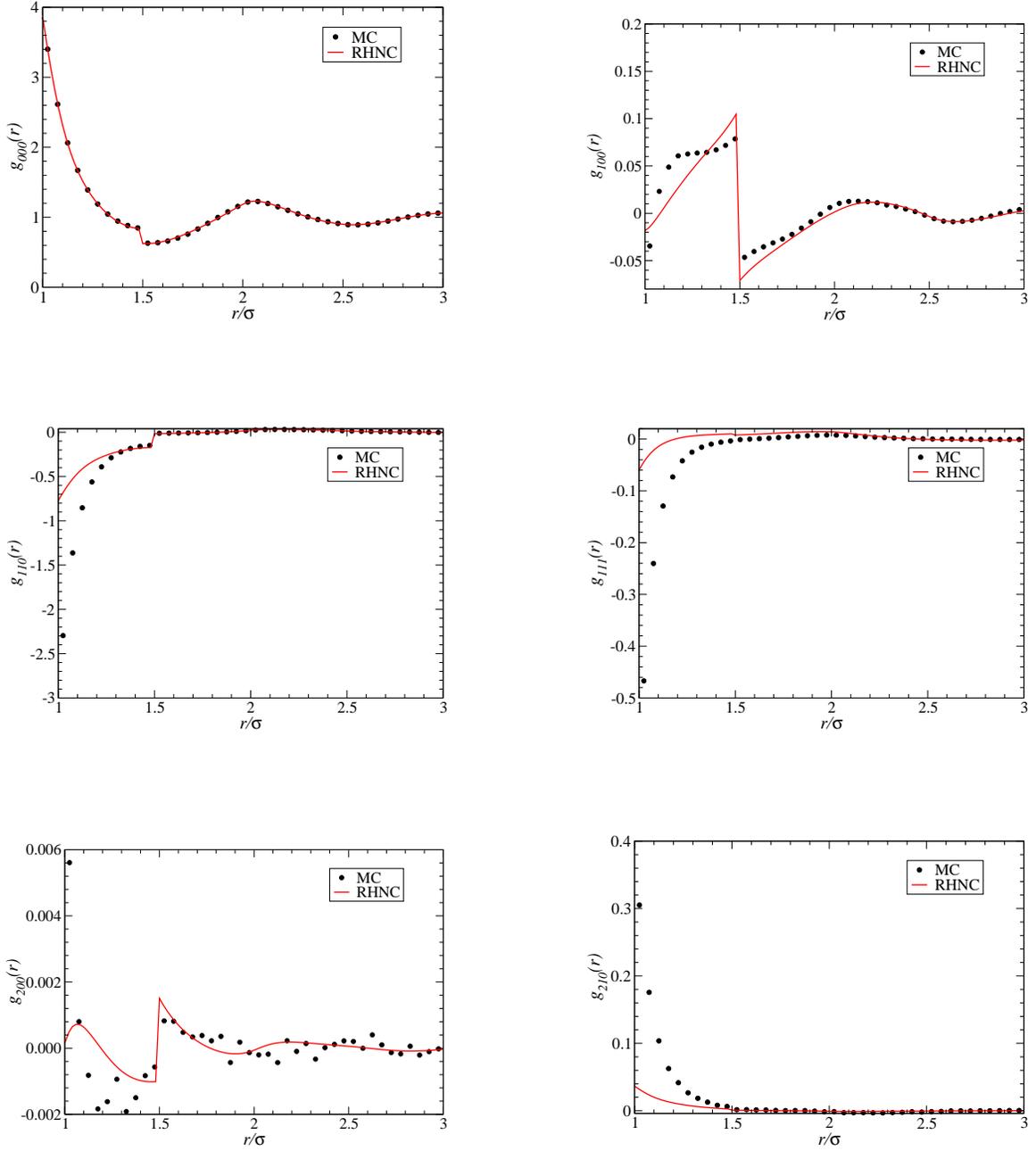}
\caption{The $g_{l_1l_2m}(r)$ rotational molecular-frame components of the pair distribution function for several triplets. The system is the Janus case (coverage $\chi=0.5$, $\lambda=1.5$ and $T^{*}=1$). Notice the change of scale in each panel. }
\label{fig:fig7}
\end{figure}

Data in Fig. \ref{fig:fig7} are for $\lambda=1.5, \chi=0.5, \rho=0.8$ and $T^{*}=1$.
Simulation results have been calculated according to the procedure described in Appendix \ref{app:appb}.
The first observation is that the spherically symmetric HS bridge function and the thermodynamically-consistent determination of its optimal diameter are able to bring the RHNC $g_{000}(r)$ into excellent agreement with computer simulation results. As expected, however, residual discrepancies, in some cases even qualitative, are observed in the non-spherical components, although the worst cases are also quantitatively less serious. The results of Fig. \ref{fig:fig7} are representative  
of the situation for all the cases we have investigated  at the same temperature and coverage ($\lambda=1.5, \rho=0.65$ and  $\rho=0.5$ ; 
$\lambda=1.2, \rho=0.8$). The natural conclusion of such comparisons is that if
one wants to improve the description of the overall structure it is important to go beyond spherical bridge function approximations. 

\subsection{Chemical potential vs pressure plane}
\label{subsec:chemical}
Having computed pressure and chemical potential as described  in Section  \ref{subsec:thermodynamics} , we can now move to the calculation of the coexistence curves by 
fixing a temperature and finding the two densities, $\rho^{*}_g$ of the gas and $\rho^{*}_l$ of the liquid, that
coexist at that temperature so as to yield equal pressures and chemical potentials. These are then the resolving densities
of the system of equations
\begin{eqnarray}
\label{chemical:eq4a}
P_{g} \left(T^{*},\rho^{*}_{g} \right)&=& P_{l} \left(T^{*},\rho^{*}_{l} \right), \\
\label{chemical:eq4b}
\mu_{g} \left(T^{*},\rho^{*}_{g} \right)&=& \mu_{l} \left(T^{*},\rho^{*}_{l} \right)\,\mbox{.}
\end{eqnarray} 

\begin{figure}[ht] 
   \includegraphics[width=6.0in]{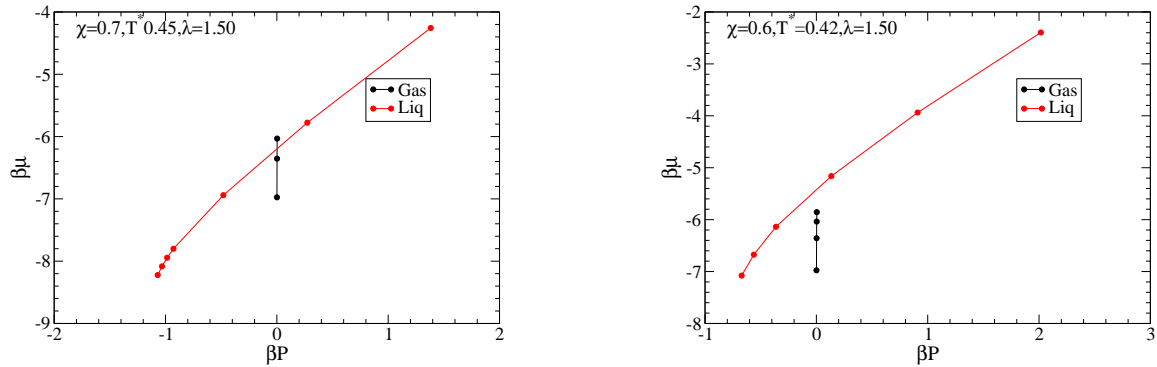}
\caption{The gas and liquid branches at fixed temperature, chemical potential, and pressure. The case 
$\chi=0.7$, $T^{*}=0.45$ is an example of real crossing; in the case 
$\chi=0.6$, $T^{*}=0.42$ the consistence condition can be obtained as safe smooth extrapolation of the two branches.} 
\label{fig:fig8}
\end{figure}

The resulting intersections are depicted in Fig. \ref{fig:fig8} for a couple of typical situations ($\chi=0.7$ and $\chi=0.6$).
Note that at the lowest coverages considered ($\chi=0.6$ and $\chi=0.5$), the crossing has to be obtained by
extrapolating the two curves. Given the improved algorithm we are using, we are inclined to attribute the crossing failure to the closure, more than to difficulties of convergence. This might also be taken as an indication of a decrease in the accuracy for the computed coexistence curves. As we will see, this turns out to be the case.

\subsection{Phase diagram}
\label{subsec:phase}
\begin{figure}[htbp]
\begin{center}
\vskip0.5cm
\includegraphics[width=10cm]{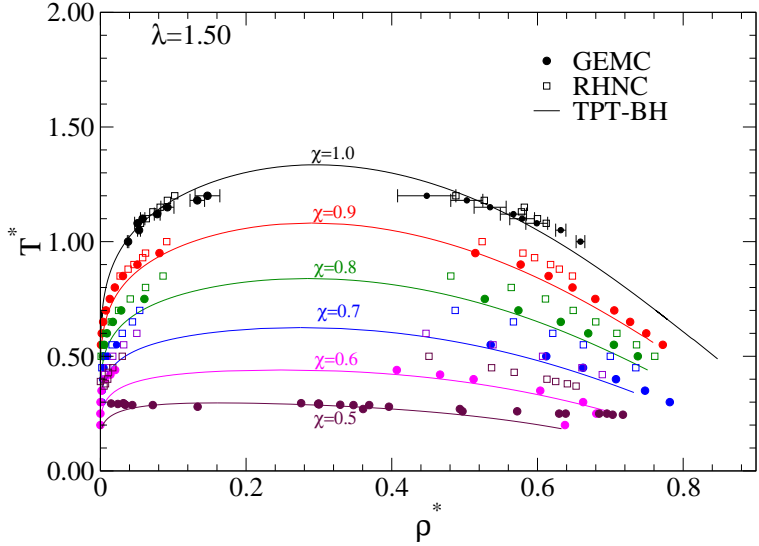} \\
\caption{The phase diagram in the temperature-density plane as a function of the coverage.
 Results reported are from GEMC simulation, RHNC integral equation theory, and TPT-BH perturbation theory.}
\label{fig:fig9}
\end{center}
\end{figure}

As discussed above, the system of Eqs. (\ref{chemical:eq4a}) and (\ref{chemical:eq4b}) provides the coexisting densities $\rho^{*}_{g}$  of the gas phase and $\rho^{*}_{l}$ of the liquid phase at a fixed temperature $T^{*}$. This allows the calculation of the full phase diagram in the temperature-density plane
as a function of the coverage $\chi$. The results are displayed in Fig. \ref{fig:fig9}, where those from RHNC integral equation theory are contrasted with results from Gibbs Ensemble Monte Carlo (GEMC) simulations and TPT-BH.

At first sight, the performances of both approximate approaches appear able to capture the main qualitative trends of the numerical simulations, given the
well-known shortcomings of each. Both approaches give fairly consistent gas curves that are relatively close to those from numerical simulations, although
this works better for larger than smaller coverages. For the liquid branch, however, the accuracy appears to be much less satisfactory, although TPT-BH appears to be able to follow the coverage dependence more closely than RHNC. Both approaches, however, fall short in the Janus limit ($\chi=0.5$), where the re-entrant phase diagram is found. 

\begin{figure}[htbp] 
  \centering
   \includegraphics[width=6.0in]{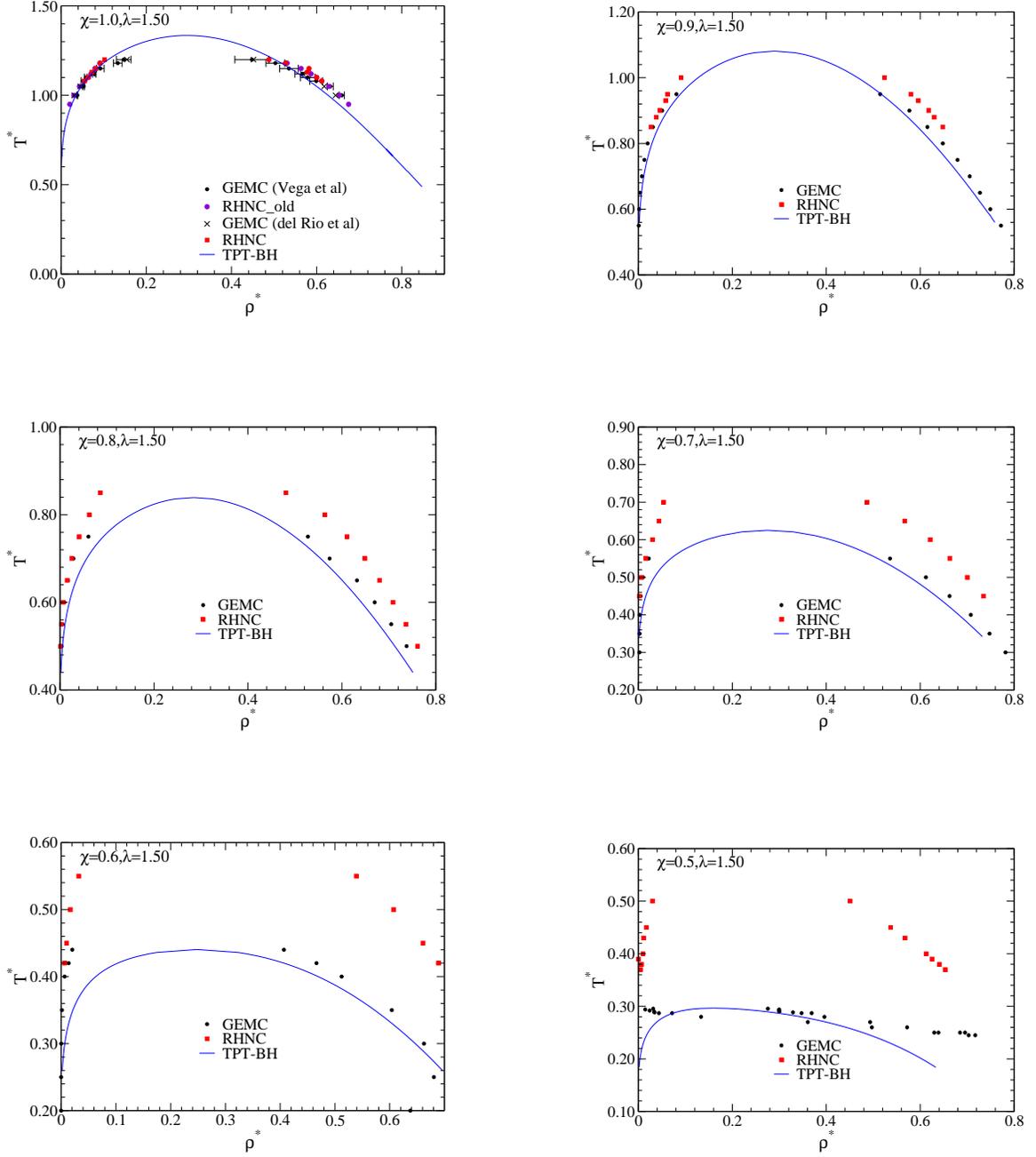} 
 \caption{The phase diagram in the temperature-density plane for various coverages: 
$\chi=1.0, 0.9,0.8,0.7,0.6,0.5$. Results reported are from GEMC simulation, RHNC integral equation theory, and TPT-BH perturbation theory.}
\label{fig:fig10}
\end{figure}

A closer look at each phase diagram, however, reveals the specific deficiencies of both RHNC and TPT-BH approaches. This is shown in Fig. \ref{fig:fig10}, where a single phase diagram in the temperature-density plane is displayed at each coverage, from a full square-well ($\chi=1.0$) potential to the Janus fluid ($\chi=0.5$) in the left-right/top-down order of decreasing coverage.
Consider the square-well $\chi=1.0$ case first. In this case, the results of numerical simulations were obtained from Vega et al. \cite{Vega92} and del R{\' i}o et al., \cite{delRio02} while the the RHNC results are based on a Newton-Raphson scheme that was pushed a little farther than a previous calculation, 
\cite{Giacometti09b} with slightly improved performance. For all other cases, the hybrid Newton-Raphson/Picard scheme previously described
was followed, allowing lower temperatures and hence lower coverages to be reached compared to the pure Picard calculation used in Ref. \onlinecite{Giacometti09a}. The TPT-BH calculations are also a refinement of those reported in Ref. \onlinecite{Gogelein12}, with little or no variation.

Within this more detailed view, the weaknesses of each approach are clearly visible. The accuracy of the RHNC approach clearly degrades as the coverage decreases, not so much by virtue of the lower temperatures involved but rather due to the intrinsic shortcoming of the spherically symmetric reference system used here for the RHNC bridge function, which becomes more and more problematic as the coverage decreases. A comparison with similar results obtained in the more isotropic two-patch case, \cite{Giacometti10} where the accuracy was much greater even with the original algorithm, strongly supports this inference.

The performance of the TPT-BH perturbation theory is based on an almost opposite scenario. As apparent from Fig. \ref{fig:fig10}, TPT-BH appears to be able to follow, albeit with some inaccuracy, the decreasing trend in terms of the coverage. On the other hand, it should be clearly emphasized that the approximation involved (see Eq. (\ref{tpt:eq2})) is independent of the way the attractive part is distributed on the surface. Notably, the prediction of  TPT-BH would be identical in the two-patches case, whereas numerical simulations indicate a significant quantitative difference in the binodal of the one-patch and the two-patch cases. A final word of caution is in order. The very good quality of perturbation theory for the one-patch case, reported in Fig. \ref{fig:fig10}, is not uniform at different values of the model parameters. For example, in the experimentally more interesting case of $\lambda=1.2$, we find significantly poorer performances of TPT-BH with respect to RHNC in reproducing coexistence curves. At the present level of investigations, a combined use of both techniques could be used to extract some first approximate information about the location of liquid-vapor coexistence.

\section{Conclusions and outlook}
\label{sec:conclusions}
In this paper we have studied the Kern-Frenkel potential with a single patch, extending to lower coverages previous work \cite{Giacometti09a} on RHNC integral equation theory. For this purpose, we implemented an improved Newton-Raphson algorithm that provides a much more stable convergence scheme at low temperatures and allowed us to decrease coverage from square-well ($\chi=1.0$) to the Janus limit ($\chi=0.5$). 

We found that, as the coverage decreases, the accuracy of RHNC
integral equation theory gradually deteriorates and we argued that this is to be attributed to the choice of the HS reference bridge function as an approximation to the real {\em anisotropic} bridge function. The contrast with the much better accuracy previously found in the two-patch calculations \cite{Giacometti10}  using the same approximation indeed strongly suggests that the origin of this shortcoming in the one-patch case stems from the highly anisotropic form of the one-patch Kern-Frenkel potential that is hardly approximated by any spherically symmetric reference model.
A second aim of our study has been a direct assessment of the pros and cons of RHNC integral equation theory compared with TPT-BH thermodynamic perturbation theory.
We found TPT-BH to be superior, for the case of $\lambda = 1.5$, in terms of its ability to predict the approximate location of the coexistence lines, its accuracy
not degrading with decreasing coverage and temperature and always at a very small computational cost. 
However, preliminary calculations for the case $\lambda = 1.2$ seem to indicate that, with decreasing range of the attractive well, TPT-BH results deteriorate faster than RHNC. Future investigations in this direction, now made possible by the improved integral equation algorithm presented in this paper, will be necessary to assess this point.
 
\begin{acknowledgments}
A.G.,F.S. and G.P. acknowledge financial support by PRIN-COFIN 2010-2011 (contract 2010LKE4CC). S.F. was financially supported by the Austrian Research Fund (FWF) under  Project P23910-N16.
\end{acknowledgments}

\appendix
\section{Explicit calculations of some rotational invariant coefficients}
\label{app:appa}
Here we provide examples of the direct calculations for rotational invariant coefficients as described in Sec. \ref{subsec:rotational}.
The general expansion of $g(12)$ for a linear molecule in an arbitrary frame was given in Eqs. (\ref{ang:eq1}) and (\ref{ang:eq2}),
where $\psi^{l_1 l_2 l}$ are the rotational invariants. In particular, we here consider explicitly the following two representative cases,
\begin{eqnarray}
\label{app:eq2}
\psi^{110}\left(\omega_1,\omega_2,\Omega \right)=\Delta\left(\omega_1,\omega_2,\Omega\right), \\
\psi^{112}\left(\omega_1,\omega_2,\Omega \right)=D\left(\omega_1,\omega_2,\Omega\right),
\end{eqnarray}
where $\Delta(\omega_1,\omega_2,\Omega)$ and $D(\omega_1,\omega_2,\Omega)$ are defined in Eq. (\ref{rotational:eq2}). The aim of this Appendix is to
compute the corresponding coefficients, as given in Eq. (\ref{rotational:eq1}).
\subsection{Calculation of $h^{l_{1}l_{2}l}(r)$}
\label{subsec:h110}
The configurational partition function for this problem is
\begin{eqnarray}
\label{app:eq3}
Z_{N} = \int \left[\prod_{l=1}^N d \mathbf{r}_l d \omega_l \right] e^{-\beta \sum_{l<m} \Phi\left(\mathbf{r}_{lm},\hat{\mathbf{n}}_l, \hat{\mathbf{n}}_m \right)}.
\end{eqnarray}

Using Eqs. (\ref{ang:eq1}) and (\ref{app:eq2}), we have
\begin{eqnarray}
\label{appa:eq1}
\left \langle \sum_{i < j} \delta\left(r-r_{ij}\right) \Delta\left(\omega_1,\omega_2,\Omega\right) \right \rangle&=&
\left \langle \sum_{i < j} \delta\left(r-r_{ij}\right)  \psi^{110}\left(\omega_1,\omega_2,\Omega \right) \right \rangle \\ \nonumber
&=& \frac{1}{Z_N} \int \left[\prod_{l=1}^N d \mathbf{r}_l d \omega_l \right] \left \langle \sum_{i < j} \delta\left(r-r_{ij}\right)  \psi^{110}\left(\omega_1,\omega_2,\omega \right) \right \rangle
e^{-\beta \sum_{l<m} \Phi\left(\mathbf{r}_{lm},\hat{\mathbf{n}}_l, \hat{\mathbf{n}}_m  \right)} \\ \nonumber
&=& \int d\mathbf{r}_{1} d\mathbf{r}_{2} \int d\omega_{1} d \omega_{2} \delta\left(r-r_{12}\right) \psi^{110}\left(\omega_1,\omega_2,\Omega\right)
\rho\left(\mathbf{r}_{12},\omega_1,\omega_2, \right),
\end{eqnarray}
where
\begin{eqnarray}
\label{appa:eq2}
\rho\left(\mathbf{r}_{12},\omega_{1},\omega_{2} \right)=\frac{N\left(N-1\right)}{2} \frac{1}{Z_N} \int \left[  d\mathbf{r}_{3} d \omega_{3} \ldots
 d\mathbf{r}_{N} d \omega_{N} \right] e^{-\beta \sum_{l<m} \Phi\left(\mathbf{r}_{lm},\hat{\mathbf{n}}_l, \hat{\mathbf{n}}_m \right)} = g\left(\mathbf{r}_{12},\omega_{1},\omega_{2}
\right) \frac{\rho^2}{\left(4 \pi\right)^2}.
\end{eqnarray}
Using then Eq. (\ref{ang:eq1}), along with the results
\begin{eqnarray}
\label{appa:eq3}
\left \langle \psi^{l_1 l_2 l}\left(\omega_1,\omega_2,\Omega\right) \psi^{l_1^{\prime} l_2^{\prime} l^{\prime}}\left(\omega_1,\omega_2,\Omega\right)
\right \rangle_{\omega_1,\omega_2,\Omega} &=& \delta_{l_1 l_1^{\prime}} \delta_{l_2 l_2^{\prime}}  \delta_{l_3 l_3^{\prime}}
 \left \langle \left[\psi^{l_1 l_2 l}\left(\omega_1,\omega_2,\Omega\right) \right]^2\right \rangle_{\omega_1,\omega_2,\Omega}
\end{eqnarray}
and 
\begin{eqnarray}
\label{appa:eq4}
\left \langle \left[\psi^{110}\left(\omega_1,\omega_2,\Omega\right) \right]^2\right \rangle_{\omega_1,\omega_2,\Omega}&=& \frac{1}{3},
\end{eqnarray}
we find, from Eq. (\ref{appa:eq2}),
\begin{eqnarray}
\label{appa:eq5}
\left \langle \sum_{i < j} \delta\left(r-r_{ij}\right) \Delta\left(\omega_1,\omega_2,\Omega\right) \right \rangle
&=&4 \pi \rho N r^2 
\sum_{l_1,l_2,l} g^{l_{1}l_{2}l}\left(r\right) \delta_{l_1 1} \delta_{l_2 1}  \delta_{l_3 0} 
\left \langle \left[\psi^{110}\left(\omega_1,\omega_2,\Omega\right) \right]^2\right \rangle_{\omega_1,\omega_2,\Omega} \\ \nonumber
&=& \frac{4 \pi}{3} \rho N r^2 g^{110}\left(r\right),
\end{eqnarray}
so that
\begin{eqnarray}
\label{appa:eq6}
g^{110}\left(r\right)=h^{110}\left(r\right) &=& \frac{3}{4 \pi \rho N r^2} 
\left \langle \sum_{i < j} \delta\left(r-r_{ij}\right) \Delta\left(\omega_1,\omega_2,\Omega\right) \right \rangle,
\end{eqnarray}
in agreement with Eq. (\ref{ang:eq2}) and Ref. \onlinecite{Weis93}.
\subsection{Calculation of $h^{112}(r)$}
\label{subsec:h112}
A similar calculation leads to the expression for $h^{112}(r)$: 
\begin{eqnarray}
\label{appa:eq7}
\left \langle \sum_{i < j} \delta\left(r-r_{ij}\right) D\left(\omega_1,\omega_2,\Omega\right) \right \rangle&=&
\left \langle \sum_{i < j} \delta\left(r-r_{ij}\right)  \psi^{112}\left(\omega_1,\omega_2,\Omega \right) \right \rangle \\ \nonumber
&=& \int d\mathbf{r}_{1} d\mathbf{r}_{2} \int d\omega_{1} d \omega_{2} \delta\left(r-r_{12}\right) \psi^{112}\left(\omega_1,\omega_2,\Omega\right)
\rho\left(\mathbf{r}_{12},\omega_{1},\omega_{2} \right).
\end{eqnarray}
Using Eqs. (\ref{appa:eq2}) and (\ref{appa:eq3}), along with the result
\begin{eqnarray}
\label{appa:eq8}
\left \langle \left[\psi^{112}\left(\omega_1,\omega_2,\Omega\right) \right]^2\right \rangle_{\omega_1,\omega_2,\Omega}&=& \frac{2}{3},
\end{eqnarray}
one finds
\begin{eqnarray}
\label{appa:eq9}
g^{112}\left(r\right)=h^{112}\left(r\right) &=& \frac{3}{8 \pi \rho N r^2} 
\left \langle \sum_{i < j} \delta\left(r-r_{ij}\right) D\left(\omega_1,\omega_2,\Omega\right) \right \rangle,
\end{eqnarray}
again in agreement with Eq. (\ref{ang:eq2}) and Ref. \onlinecite{Weis93}.
\section{MC calculation of the molecular reference coefficients
$g_{l_1l_2m}(r)$}
\label{app:appb}
The molecular reference coefficients
$g_{l_1l_2m}(r)$ are related to the angular dependent pair distribution function $g(r,\omega_1,\omega_2)$ by 
\begin{eqnarray}
\label{appb:eq1}
g_{l_{1} l_{2} m} \left(r\right)&=&\frac{1}{4\pi} \int d\omega_1d\omega_2\, g\left(r,\omega_1,\omega_2 \right) 
Y_{l_{1}m}^*\left(\omega_1\right)
Y_{l_{2}\bar{m}}^*\left(\omega_2\right) \nonumber \\
&=&
4 \pi \left \langle g\left(r,\omega_1,\omega_2 \right) 
Y_{l_{1}m}^*\left(\omega_1\right)
Y_{l_{2}\bar{m}}^*\left(\omega_2\right) \right \rangle_{\omega_{1},\omega_{2}}. 
\end{eqnarray}
By multiplying and dividing Eq. (\ref{appb:eq1}) by $g_{000}(r)=\langle g(r,\omega_1,\omega_2) \rangle_{\omega_1,\omega_2}$, it can be cast in
the following form:
\begin{eqnarray}
\label{appb:eq2}
g_{l_1 l_2 m}(r) =  4 \pi g_{000}(r) 
\frac{
\left \langle g(r,\omega_1,\omega_2) Y^{*}_{l_1m}\left(\omega_1\right) Y^{*}_{l_2\bar{m}}\left(\omega_2\right)
\right \rangle_{\omega_1,\omega_2}
}{\left \langle g(r,\omega_1,\omega_2) \right \rangle_{\omega_1,\omega_2}}.
\end{eqnarray}
Upon introducing the new average
\begin{eqnarray}
\label{appb:eq3}
\left \langle \ldots \right \rangle_{r} &\equiv& \frac{\left \langle g(r,\omega_1,\omega_2) \ldots \right \rangle_{\omega_1,\omega_2}}
{\left \langle g(r,\omega_1,\omega_2) \right \rangle_{\omega_1,\omega_2}},
\end{eqnarray}
where the subscript $r$ of the average means that it is restricted to particle centers at separation $r$, we find
\begin{eqnarray}
\label{appb:eq4}
g_{l_1 l_2 m}(r) =  4 \pi g_{000}(r) \left \langle Y^{*}_{l_1m}\left(\omega_1\right) Y^{*}_{l_2\bar{m}}\left(\omega_2\right) \right \rangle_r.
\end{eqnarray}

\begin{thebibliography}{99}

\bibitem{Walther09} A. Walther and A. H. E. M\"uller, Soft Matter \textbf{4}, 663 (2008).

\bibitem{Pawar10} A. B. Pawar and I. Kretzschmar, Macromol. Rapid Commun. \textbf{31}, 150 (2010).

\bibitem{Glotzer04} S. C. Glotzer, Science \textbf{306}, 419 (2004).

\bibitem{Glotzer07} S. C. Glotzer and M. J. Solomon, Nature Mater. \textbf{6}, 557 (2007).

\bibitem{Whitesides02} G. M. Whitesides and M. Boncheva, Proc. Natl. Acad. Sci. \textbf{99}, 4769 (2002); G. M. Whitesides and 
B. Grzybowski, Science \textbf{295}, 2418 (2002)..

\bibitem{Sciortino09} F. Sciortino, A. Giacometti, and G. Pastore, Phys. Rev. Lett. \textbf{103}, 237801 (2009).

\bibitem{Sciortino10} F. Sciortino, A. Giacometti, and G. Pastore, Phys. Chem. Chem. Phys. \textbf{12}, 11869 (2010).

\bibitem{Gray84} C. G. Gray and K. E. Gubbins, \textit{Theory of Molecular Fluids. Volume 1: Fundamentals} (Clarendon Press, Oxford, 1984).

\bibitem{Hansen86} J. P. Hansen and I. R. McDonald, \textit{Theory of Simple Liquids} (Academic, New York, 1986).

\bibitem{Kern03} N. Kern and D. Frenkel, J. Chem. Phys. \textbf{118}, 9882 (2003).

\bibitem{Chapman88} W. G. Chapman, G. Jackson, and K. E. Gubbins, Mol. Phys. \textbf{65}, 1057 (1988).

\bibitem{Giacometti09a} A. Giacometti, F. Lado, J. Largo, G. Pastore, and F. Sciortino, J. Chem. Phys. \textbf{131}, 174114 (2009).

\bibitem{Giacometti10} A. Giacometti, F. Lado, J. Largo, G. Pastore, and F. Sciortino, J. Chem. Phys. \textbf{132}, 174110 (2010).

\bibitem{Giacometti09b} A. Giacometti, G. Pastore, and F. Lado, Mol. Phys. \textbf{107}, 555 (2009).

\bibitem{Gogelein12} C. G\"ogelein, F. Romano, F. Sciortino, and A. Giacometti, J. Chem. Phys. \textbf{136}, 094512 (2012).

\bibitem{Lado73} F. Lado, Phys. Rev. A \textbf{8}, 2548 (1973).

\bibitem{Rosenfeld79} Y. Rosenfeld and N. W. Ashcroft, Phys. Rev. A \textbf{20}, 1208 (1979).

\bibitem{Barker67} J. A. Barker and D. Henderson, J. Chem. Phys. \textbf{47}, 2856 (1967).

\bibitem{Barker76} J. A. Barker and D. Henderson, Rev. Mod. Phys. \textbf{48}, 587 (1976).

\bibitem{Lado82a} F. Lado, Mol. Phys. \textbf{47}, 283 (1982).

\bibitem{Lado82b} F. Lado, Mol. Phys. \textbf{47}, 299 (1982).

\bibitem{note1} The present formulation of the potential follows Ref. \onlinecite{Gogelein12} and formally differs from that given
in Refs. \onlinecite{Kern03,Giacometti09a,Giacometti10} that, strictly speaking, is slightly inconsistent. The actual form used in all numerical calculations
was however always correct.

\bibitem{Green} M. S. Green, J. Chem. Phys. {\bf 33}, 1403 (1960); M. Klein and M. S. Green, {\em ibid.} {\bf 39}, 1367 (1963). Green classified the diagrams in the density expansion of $g(r)e^{\beta \phi(r)}$ for a spherically-symmetric potential $\phi(r)$ by analogy with electric circuits as ``series,'' ``parallel,'' or ``bridge,'' the last because of the resemblance of its first diagram to a Wheatstone bridge. The ``parallel'' diagrams can be summed in direct space and disappear. The name ``series'' for $\gamma(r)=h(r)-c(r)$ is nowadays seldom used, but the ``bridge'' name incongruously lives on.

\bibitem{Lado82} F. Lado, Phys. Lett. A \textbf{89}, 196 (1982). The multiplicative constants $\rho$ and $\sigma_0$ in Eq.(\ref{rhnc:eq13}) can obviously be eliminated from this equation. But in numerical calculations, the right-hand-side is never zero but rather a number that is ``small enough.'' Writing the left-hand-side in dimensionless form makes the standard of ``small enough'' more consistent across thermodynamic states.

\bibitem{Verlet72} L. Verlet and J. J. Weis, Phys. Rev. A \textbf{5}, 939 (1972).

\bibitem{Henderson75} D. Henderson and E. W. Grundke, J. Chem. Phys. \textbf{63}, 601 (1975).

\bibitem{Broyles60} A. A. Broyles, J. Chem. Phys. \textbf{33}, 456 (1960).

\bibitem{Ng74} See Appendix in K. C. Ng, J. Chem. Phys. \textbf{61}, 2680 (1974). 

\bibitem{Gillan79} M. J. Gillan, Mol. Phys. \textbf{38}, 1781 (1979).

\bibitem{Labik85} S. Lab\'{i}k, A. Malijevsk\'{y}, and P. Vo\v{n}ka, Mol. Phys. \textbf{56}, 709 (1985).

\bibitem{Lomba89} E. Lomba, Mol. Phys. \textbf{68}, 87 (1989).

\bibitem{lado67} F. Lado, J. Chem. Phys. \textbf{47}, 4828 (1967).

\bibitem{Henderson71} D. Henderson and J. A. Barker, \textit{Physical Chemistry, an Advanced Treatise} (Academic Press, New York, 1971), Vol. VIIIA, p. 377.

\bibitem{CS69} N. F. Carnahan and K. E. Starling, J. Chem. Phys. \textbf{51}, 635 (1969).


\bibitem{Martinez-Haya03} B. Mart\'{\i}nez-Haya, A. Cuetos, and S. Lago, Phys. Rev. E \textbf{67}, 051201 (2003); see also
I. Nezbeda and T. Boublik, Czech. J. Phys. \textbf{28}, 353 (1978).

\bibitem{Stell81} See Appendix B in G. Stell, G. N. Patey, and J. S. H{\o}ye, Adv. Chem. Phys. \textbf{48}, 183 (1981). 

\bibitem{Vega92} L. Vega, E. de Miguel, L. F. Rull, G. Jackson, and I. A. McLure, J. Chem. Phys. \textbf{96}, 2296 (1992).

\bibitem{delRio02} F. del R\'{i}o, E. \'{A}valos, R. Esp\'{i}ndola, L. F. Rull, G. Jackson, S. Lago, Mol. Phys. \textbf{100}, 2531 (2002).

\bibitem{Weis93} J. J. Weis and D. Levesque, Phys. Rev. E \textbf{48}, 3728 (1993).



\end{thebibliography}



\end{document}